\documentclass[aps,preprint,amsmath,amssymb,nofootinbib]{revtex4}
\pdfoutput=1
\usepackage[latin1]{inputenc}
\usepackage{bbm}
\usepackage{bm}
\usepackage{graphicx}
\usepackage{epsfig}
\usepackage{subfigure}
\usepackage{latexsym}
\usepackage{amsmath}
\usepackage{amsfonts}
\usepackage{amssymb}
\usepackage{wasysym}
\usepackage{color}

\def\tr{\mathrm{Tr}}

\newcommand{\be}{\begin{equation}}
\newcommand{\ee}{\end{equation}}

\newcommand{\bea}{\begin{eqnarray}}
\newcommand{\eea}{\end{eqnarray}}

\begin{document}


\title{ {\Large Ultraviolet Complete Technicolor \\ and Higgs Physics at LHC}
} 
\author{Matti Antola\footnote{Currently at Valo Research and Trading, Helsinki, Finland}
}
\email{antola.matti@gmail.com}
\affiliation{Helsinki Institute of Physics, P.O.Box 64, FI-000140, Univ. of Helsinki, Finland} 
\author{Stefano Di Chiara}
\email{dichiara@helsinki.fi}
\affiliation{Helsinki Institute of Physics, P.O.Box 64, FI-000140, Univ. of Helsinki, Finland}
\author{Kimmo Tuominen}
\email{kimmo.i.tuominen@jyu.fi}
\affiliation{Department of Physics, P.O.Box 35, FI-40014, Univ. of Jyv\"askyl\"a, Finland}
\affiliation{Helsinki Institute of Physics, P.O.Box 64, FI-000140, Univ. of Helsinki, Finland}

\begin{abstract}
We construct a Technicolor model which provides masses for the electroweak gauge bosons and for all the Standard Model matter fields. Starting from an ultraviolet complete supersymmetric technicolor, we propose a scenario where all elementary scalars, gauginos, and higgsinos are decoupled at an energy scale substantially higher than the electroweak scale, therefore avoiding the little hierarchy problem of the minimal supersymmetric standard model. The resulting low energy theory has an SU(3) global symmetry whose breaking to SO(3) leads to electroweak symmetry breaking. We study in detail the phenomenology of this theory and demonstrate that it reproduces  the present LHC data at the same level
of precision as the Standard Model itself.  

\end{abstract}
\maketitle
 
\section{Introduction}
\label{Intro}
The data collected at the LHC experiments during the 7 and 8 TeV runs, with the epochal discovery of the Higgs boson \cite{Aad:2012tfa,Chatrchyan:2012ufa} and the measurement of its couplings 
\cite{ATLAS-CONF-2013-034,CMS-PAS-HIG-13-005}, seem to have provided the experimental 
verification of the Standard Model (SM) in its entirety. 
Compatibility of the present LHC data with the SM predictions does not imply that no beyond SM
physics exists. Rather, there are several reasons to expect the contrary: The SM Higgs does not explain why the electroweak scale exists, but merely parametrizes its consequences. The masses of the SM fermions are not explained but remain as parameters fitted to their measured values; in particular, SM provides no insight to the flavor patterns or why there exists three generations of matter fields. There is no dark matter candidate in SM nor an explanation for the matter-antimatter asymmetry.

What the LHC data clearly indicates, is that the new physics coupled with the electroweak SM currents must have a typical scale of the order of a TeV or higher. One possibility is that the new physics scale is much above the terascale, and then naturality as a model building paradigm should be reinterpreted \cite{Heikinheimo:2013fta,Heikinheimo:2013xua}. 
If the new physics scale on the other hand is near the terascale, then the present LHC data should provide 
hints of a new spectrum awaiting discovery in the future runs at the LHC. In this paper we investigate
a model framework falling into the latter category.

In Technicolor (TC) theories  \cite{Weinberg:1975gm,Susskind:1978ms} the new physics scale
is naturally of the order of a TeV. The electroweak (EW) symmetry is broken by a new strong interaction which generates a fermion 
condensate, in a way analogous to QCD but with a technipion decay constant approximately equal to the EW 
scale. The absence of light elementary scalars in TC automatically solves the fine-tuning problem traditionally associated with such massive fields without any protective symmetries. 
Gauge theories whose beta function stays small over a large range of energies constitute a popular class of TC models called walking TC \cite{Holdom:1984sk,Yamawaki:1985zg}. The lightest scalar in walking TC is the pseudo Nambu-Goldstone boson associated with the weakly broken conformal symmetry, and is therefore expected to be lighter than the other composite states \cite{Dietrich:2005jn,Elander:2009pk,Grinstein:2011dq,Antipin:2011aa}.
Recently it has been shown that a further suppression of the naive prediction for the composite Higgs mass, obtained by scaling up the mass of the $\sigma$ state of QCD \cite{Hill:2002ap}, arises from the loop corrections from the heavy SM states \cite{Foadi:2012bb}. These alone might explain the measured value, 125 GeV, of the Higgs mass.

To explain the observed mass patterns of the known matter fields within the TC framework one needs to 
couple TC fields with the matter fields of the SM. A well-known approach is extended TC (ETC)
\cite{Dimopoulos:1979es,Eichten:1979ah}, in which the technicolored femions (technifermions) and the SM fermions are embedded into a representation of a larger gauge group containing the SM and TC gauge groups. 
If the ETC gauge group breaks sequentially, such ETC model may explain the observed mass hierarchies of the SM fermions \cite{Appelquist:1993sg,Appelquist:2003hn}. 

Another framework for extending TC with couplings to SM matter fields is bosonic TC (BTC).
In BTC the couplings between SM fermions and composite scalars are introduced by assuming that both SM fermions and technifermions have Yukawa couplings to elementary heavy Higgs scalars 
\cite{Simmons:1988fu,Kagan:1991gh,Carone:1992rh,Carone:1994mx,Antola:2009wq}. To control fine-tuning at  scales above the mass scales of the elementary scalars 
one can introduce supersymmetry (SUSY) \cite{Dine:1981za,Dobrescu:1995gz}.

Recently we constructed a model along these lines and called it Minimal Super Conformal Technicolor (MSCT) \cite{Antola:2010nt,Antola:2011bx}. In MSCT the low energy effective theory is Minimal Walking Technicolor (MWT)  
\cite{Sannino:2004qp,Dietrich:2005jn,Foadi:2007ue} whose global symmetry is SU$(4)$. 
In this paper we describe another possibility within this model framework. We show that even though the particle content of the model is large, the low energy mass spectrum required by phenomenological viability can be simple: All the elementary scalars and gauginos are assumed heavy and integrated out. At the TeV scale this leaves, besides the SM fermions and gauge bosons, three Weyl technifermions, a technigluon, and a heavy  fourth generation EW doublet of leptons. The TC sector of the mesoscopic Lagrangian obtained in this case features an SU$(3)$ global symmetry; due to its global symmetry and the resemblance to MWT we call this model SU$(3)$ Minimal Walking Technicolor (3MWT).

The 3MWT offers a concrete example of how the flavor physics sector can have a large effect on the low energy phenomenology, when compared to the underlying TC model without the coupling to the flavor sector. The four fermion interactions responsible for this deformation in our low energy theory result from attractive Yukawa interactions. This can be contrasted with the traditional ETC approach where the corresponding operators arise from gauge interactions which can be attractive or repulsive. 

In Section~\ref{microL} we briefly introduce MSCT as an ultraviolet (UV) complete TC theory and review its main features \cite{Antola:2011bx}.  In Section~\ref{eff} we derive the mesoscopic Lagrangian of 3MWT, by integrating out the heavy states below the SUSY breaking scale, $m_{\rm SUSY}$, 
but above the dynamical scale of TC,  $\Lambda_{\rm TC}$. 
In Section~\ref{pure} we determine the effective Lagrangian at scales below 
$\Lambda_{\rm TC}$ in terms of SM fields, composite scalars (including the light Higgs boson), and 
composite vector degrees of freedom. 
In Section~\ref{pheno} we study the phenomenology of the model. Then, in Section~\ref{exp} we perform a goodness of fit analysis for 3MWT based on the current collider data (from both LHC and Tevatron) for Higgs physics, as well as the precision EW parameters $S$ and $T$ \cite{Peskin:1990zt,Peskin:1991sw}. We show that the data favors 3MWT over the SM. In Section~\ref{conc} we present our conclusions.

\section{MSCT: a UV Complete Technicolor}
\label{microL}

In MSCT the flavor extension (generating the SM fermion masses) corresponds to adding two Higgs 
scalar EW doublets with opposite hypercharge to MWT and supersymmetrizing the whole theory. The 
resulting model is gauge invariant under SU$(2)_{\rm TC}\times$SU$(3)_c\times$SU$(2)_L\times$U$(1)_Y$ 
and can 
be viewed as the Minimal Supersymmetric Standard Model (MSSM) extended by a fourth lepton family and 
the TC sector which, due to supersymmetrization, is $\cal N$=4 Super Yang Mills (4SYM) theory. The non-MSSM superfields appearing in MSCT and their quantum numbers 
are given in Table~\ref{MSCTsuperfields}. 
\begin{table}
\begin{tabular}{|c|c|c|c|}
\hline 
Superfield  & SU$(2)_{\rm TC}$  & SU$(2)_{\text{L}}$  & U$(1)_{\text{Y}}$\tabularnewline
\hline 
\hline 
$\Phi_{L}$  & Adj  & $\square$  & 1/2\tabularnewline
$\Phi_{3}$  & Adj  & 1  & -1\tabularnewline
$V$  & Adj  & 1  & 0\tabularnewline
$\Lambda_{L}$  & 1  & $\square$  & -3/2\tabularnewline
$N$  & 1  & 1  & 1\tabularnewline
$E$  & 1  & 1  & 2\tabularnewline
\hline 
\end{tabular}\caption{\label{MSCTsuperfields} Non-MSSM superfield content of MSCT.
Here Adj and $\square$ denote the adjoint and fundamental representations,
respectively. None of the fields above is charged under SU(3)$_{c}$.}
\end{table}

The renormalizable lepton and baryon number conserving superpotential
for MSCT is 
\begin{equation}
P=P_{\rm MSSM}+P_{\rm TC},\label{spot-1}
\end{equation}
 where $P_{\rm MSSM}$ is the MSSM superpotential, and 
\begin{equation}
P_{\rm TC}=-\frac{g_{\rm TC}}{\sqrt{2}}\epsilon^{abc}\Phi_{L}^{a}\cdot\Phi_{L}^{b}\Phi_{3}^{c}+y_{U}\Phi_L^{a}\cdot H_{u}\Phi_{3}^{a}+y_{N}\Lambda_L \cdot H_{u}N+y_{E}\Lambda_L\cdot H_{d}E+y_{R}E\Phi_{3}^{a}\Phi_{3}^{a}\ ,\label{spmwt-1}
\end{equation}
with $H_u$ ($H_d$) denoting the $Y=+1/2\left(-1/2\right)$ Higgs superfield.
The dot ($\cdot$) indicates a contraction between the SU$(2)_{L}$ doublets with the antisymmetric
two-index Levi-Civita tensor $\epsilon$.

To the Lagrangian corresponding to the superpotential in Eq.~\eqref{spot-1} we add the soft SUSY breaking terms  of the MSSM as well as those corresponding to $P_{\rm TC}$, with the latter expressed by:
\begin{eqnarray}
{\cal L}^{\rm TC}_{soft} & = & -\left[a_{\rm TC}\epsilon^{abc}\hat{U}_{L}^{a}\hat{D}_{L}^{b}\hat{{U}}_{R}^{*c}
+a_{U}\hat{Q}_{L}^{a}\cdot\hat{H}_{u}\hat{{U}}_{R}^{*a}
+a_{N}\hat{\Lambda}_{L}\cdot\hat{H}_{u}\hat{{N}}_{R}^*\right.\nonumber \\
 & + & \left.a_{E}\hat{\Lambda}_{L}\cdot\hat{H}_{d}\hat{{E}}_{R}^*
+a_{R}\hat{{U}}_{R}^{*a}\hat{{U}}_{R}^{*a}\hat{{E}}^*_{R}
+\frac{1}{2}M_{D}{D}_{R}^{\dagger a}{D}_{R}^{\dagger a}+c.c.\right]-M_{Q}^{2}\hat{{Q}}_{L}^{\dagger a}\hat{Q}_{L}^{a}\nonumber \\
 & - & M_{U}^{2}\hat{{U}}_{R}^{*a}\hat{U}_{R}^{a}-M_{L}^{2}\hat{{L}}^\dagger_{L}\hat{L}_{L}-M_{N}^{2}\hat{{N}}_{R}^*\hat{N}_{R}-M_{E}^{2}\hat{{E}}^*_{R}\hat{E}_{R}.
\end{eqnarray}
We denote the scalar components of a chiral supermultiplet
with a hat.

This model constitutes our fundamental description in terms of the
elementary degrees of freedom and forces. The relevant scales of the
problem are the SUSY breaking scale, $m_{\rm SUSY}$, and the
EW scale that we identify with the low-energy strongly coupled regime
of the TC theory $\Lambda_{\rm TC}\sim4\pi v_{w}$, which for $v_{w} = 246$~GeV
implies $\Lambda_{\rm TC}\sim3$~TeV. We will assume here that the two scales
satisfy
\begin{equation}
m_{\textrm{SUSY}}>\Lambda_{\rm TC}.
\end{equation}
With this ordering the EW symmetry is broken dynamically. Furthermore, we arrange
the spectrum in the following way:
\begin{itemize} 

\item[1)] All SUSY breaking masses are taken to be of the order of $m_\textrm{SUSY}$.

\item[2)]  We will argue that in our model the $\mu$ parameter, which gives the mass of the Higgsinos, is much larger than 
$\Lambda_{\textrm{\rm TC}}$. Therefore the Higgsinos are ignored in the low energy phenomenology.

\item[3)] The remaining composite states acquire a mass, which at most is of the order of 
$\Lambda_{\rm TC}$. 

\end{itemize}

After the identification of the relevant scales and their ordering we now start systematically deriving the effective Lagrangian description towards the electroweak scale.

\section{Mesoscopic Lagrangian\label{eff}}

In this section we derive the full Lagrangian of 3MWT, including the four fermion interactions, by integrating out all elementary scalars, all gauginos, and all higgsinos.  The key point of this model is
that the technigaugino $D^\dagger_R$ is removed from low energy dynamics, and this modifies the initial technicolor theory, MWT, strongly.
Now the particle content of the TC theory is
\be Q_L^a=\left(\begin{array}{c} U^{a}_L \\D^{a}_L \end{array}\right) , \qquad U_R^a  \ ,  \qquad a=1,2,3 \ , \label{3MWTQ}\ee
where, in the Weyl basis, $U_L$ and $D_L$ are the left-handed techniup and technidown, respectively, 
and $U_R$ is the right-handed up-type techniparticle. These are the fermionic states of the supermultiplets denoted by $\Phi_L$ and $\Phi_3$ in Section \ref{microL}. 

To determine the 3MWT Lagrangian, valid between the TeV-scale and $m_{\rm SUSY}$, we now consider the Higgs Yukawa couplings in MSCT:
\begin{eqnarray}
-\mathcal{L}^{\rm MSCT}_{\text{Yukawa}} & = & \hat{H}_{u}\cdot F_{u}+\hat{H}_{d}\cdot F_{d}+\textrm{h.c.}\ ,\nonumber \\
F_{u} & = & q_{Lu}^{i}Y_{u}^{i}u_{R}^{\dagger i}+y_{U}Q_{L}U^\dagger_{R}+y_{N}L_{L}N^\dagger_{R}\ ,\nonumber \\
F_{d} & = & q_{Ld}^{i}Y_{d}^{i}d_{R}^{\dagger i}+l_{L}^{i}Y_{l}^{i}e_{R}^{\dagger i}+y_{E}L_{L}E^\dagger_{R}\ .\label{Lmic-1}
\end{eqnarray}
The fields $\hat{H}_{u}$ and $\hat{H}_{d}$ are the two Higgs scalar
doublets. The flavor index is denoted by $i=1...3$ and it is
summed over. The matrices $Y_{u}$, $Y_{d}$, and $Y_{l}$ are diagonal,
and the CKM matrix $V$ is hidden in the definitions of the vectors
\begin{equation}
q_{Lu}^{T i}=(u_{L}^{i},V^{ij}d_{L}^{j})\qquad{\rm and}\qquad q_{Ld}^{T i}=(V^{\dagger ij}u_{L}^{j},d_{L}^{i})\ .
\end{equation}

The potential of the MSSM Higgs fields is 
\begin{equation}
V_{\textrm{\rm MSSM}}=\left(m_{\textrm{SUSY}}^{2}+|\mu|^{2}\right)|\hat{H}_{u}|^{2}+\left(m_{\textrm{SUSY}}^{2}+|\mu|^{2}\right)|\hat{H}_{d}|^{2}-\left(b\hat{H}_{u}\hat{H}_{d}+\textrm{h.c.}\right)+...\label{vin-1-1}
\end{equation}
 The Higgs states are diagonalized via: 
\begin{eqnarray}
\left(\begin{array}{c}
\hat{H}_{u}\\
\hat{H}_{d}^{c}
\end{array}\right) & = & \frac{1}{\sqrt{2}}\left(\begin{array}{cc}
1 & -1\\
1 & 1
\end{array}\right)\left(\begin{array}{c}
\hat{H}_{1}\\
\hat{H}_{2}
\end{array}\right)\ ,
\end{eqnarray}
 where $\hat{H}_{d}^{c}=i\sigma^2\hat{H}_{d}^{*}$, with $\sigma^2$ being a Pauli matrix. The tree-level
masses 
$m_{1,2}^{2}=\mu^{2}+m_{\text{SUSY}}^{2}\pm b$ of $\hat{H}_{1,2}$ 
are traded for two convenient parameters $\theta$ and $m_{s}$ defined via the following relations: 
\begin{equation}
\frac{1}{2}\left(\frac{1}{m_{1}^{2}}+\frac{1}{m_{2}^{2}}\right)=\frac{c_{\theta}^{2}}{m_{s}^{2}}\quad,\quad\frac{1}{2}\left(\frac{1}{m_{1}^{2}}-\frac{1}{m_{2}^{2}}\right)=\frac{c_{\theta}s_{\theta}}{m_{s}^{2}}.
\end{equation}
 In terms of the parameters appearing in Eq.~(\ref{vin-1-1}) we
have 
\begin{equation}
m_{s}^{2}=\left(\mu^{2}+m_{\text{SUSY}}^{2}\right)\frac{(\mu^{2}+m_{\text{SUSY}}^{2})^{2}-b^{2}}{\left(\mu^{2}+m_{\text{SUSY}}^{2}\right)^{2}+b^{2}}\quad,\quad t_\theta=\frac{b}{\mu^{2}+m_{\text{SUSY}}^{2}}\ .
\label{eq:mss}
\end{equation}
In the above equations we have denoted $\sin\theta,\cos\theta,\tan\theta$ simply with  $s_\theta,c_\theta,t_\theta$.
Decoupling all the scalars, as well as the gauginos and higgsinos, leads to the following intermediate scale (or mesoscopic)
interaction Lagrangian for the fermions: 
\begin{eqnarray}
{\cal L}_{4-fermion} & = & \frac{c_{\theta}^{2}}{m_{s}^{2}}\left(F_{u}^{\dagger}F_{u}+F_{d}^{\dagger}F_{d}\right)-\frac{c_{\theta}s_{\theta}}{m_{s}^{2}}\left(F_{u}\cdot F_{d}+\textrm{h.c.}\right)\nonumber \\
 &  & +\frac{g_{\text{\rm TC}}^{2}}{m_{\textrm{SUSY}}^{2}}\epsilon_{abc}\epsilon_{cde}\eta_{i}^{\alpha a}{\eta_{j\alpha}^{b}}\eta_{i\dot{\beta}}^{\dagger d}\eta_{j}^{\dagger \dot{\beta} e}\ ,\label{4fL-2}
\end{eqnarray}
with
\be \eta_\alpha^T= \left( U_{L \alpha} , D_{L \alpha} ,  -i \sigma^2_{\alpha\beta} U^{\dagger \beta}_R\right)\ .
\label{SU(3)multiplet} \ee 
We have included the operators of mass dimension less or equal to
six. The indices $i$ and $j$ denote SU$(3)$ flavor; the first
letters of the alphabet are reserved for the adjoint SU$(2)$ technicolor
indices, while the Greek indices label the spin component. The color
indices are contracted and suppressed, while the TC indices, running
from 1 to 3, are written explicitly only in the last term. 
We suppress summed spin indices as long as it can be done consistently.

The first line in Eq.\,\eqref{4fL-2} derives from decoupling the Higgs scalars, and breaks the global SU$(3)$ symmetry, while the last term originates from the Yukawa interaction between technisquarks and techniquarks. This interaction stems from the superpotential and gauge interactions of the $\mathcal{N}=4$ sector. Because of its origin, the last term in Eq\,.\eqref{4fL-2} respects the global SU$(3)$ symmetry of the pure TC sector.


\section{Effective Lagrangian at the Electroweak Scale}\label{pure}

In this section we derive the effective Lagrangian at scales below $\Lambda_{\rm TC}$. 
The relevant degrees of freedom are the composite states associated to the strong TC interaction, 
and the form of the effective Lagrangian is constrained by requiring the effective theory to satisfy the  approximate global symmetries of the fundamental Lagrangian. In  addition to the composite scalar fields, we introduce also composite vector fields in a consistent manner.

The effective theory at the electroweak scale can be seen as composed of two parts: 
\be
{\cal L}_{eff}(M,K,K') = {\cal L}_{\rm 3MWT}(M)+\mathcal{L}_{F}(M,K,K')
\,,
\label{Leff}
\ee
where $M$ is the composite scalar matrix.  The pure technicolor theory, corresponding to the particle content in Eq.~\eqref{3MWTQ} and the effective Lagrangian $\mathcal{L}_{\rm 3MWT}$, respects the full SU$(3)$ symmetry. The explicit breaking of this global symmetry due to the flavor extension is represented by the effective Lagrangian $\mathcal{L}_{F}$ containing the spurions $K$ and $K'$, which are defined in Subsection~\ref{FeffL}. We will now define these effective Lagrangian contributions in detail.

\subsection{Technicolor Scalar Sector}\label{SeffL}
The composite scalar matrix field $M$, singlet under SU$(2)_{\rm TC}$, has minimal particle content given by the techniquark bilinears:
\be
M_{ij} \sim \eta_i^\alpha \eta_j^\beta \varepsilon_{\alpha\beta}  =\eta_i {\eta_j}\ , \quad \quad {\rm with}\quad\quad i,j=1\dots 3.
\label{M-composite}
\ee
The field $M$ transforms under the full SU(3) group according to
\be
M\rightarrow {\rm u} M {\rm u}^T \ , \qquad {\rm with} \qquad {\rm u}\in {\rm SU(3)} \ .
\ee
The effective linearly transforming SU$(3)$ invariant Lagrangian reads:
 \be
{\cal L}_{M} = \frac{1}{2}\text{Tr}\left[D_\mu M^{\dagger}D^\mu M\right]-\mathcal{V}_{M}
\ ,
\label{LMWT}
\ee
where the covariant derivative is given by
\[
D_\mu M=\partial_\mu M-i g_L\left[G_\mu M+M G_\mu^T\right]\;,
\]
with
\be\label{Gm}
 G_\mu=\tilde{W}^a_\mu\frac{\lambda^a}{2}+t_\xi B_\mu Y_M\ ,\quad a=1,2,3\ .
\ee
In the above equation $\lambda^a$ are the Gell-Mann matrices,  $t_\xi=\tan\xi$ with $\xi$ the EW mixing angle, $\tilde{W}_\mu$ and $B_\mu$ are the SM EW gauge fields, and
\be
Y_M={\textrm{diag}}\left(\frac{1}{2},\frac{1}{2}, -1\right).
\ee
The most general SU$(3)$  preserving effective potential, including only  up to dimension four operators, is
\be
\mathcal{V}_{M}=
-\frac{m^{2}}{2}\text{Tr}\left[M^{\dagger}M\right]
+\frac{\lambda}{4}\text{Tr}\left[M^{\dagger}M\right]^{2}
+\lambda'\text{Tr}\left[M^{\dagger}MM^{\dagger}M\right]
-2m'\left[\det M+\det M^{\dagger}\right]  \ ,
\label{Vsu3} 
\ee
which breaks SU$(3)$ spontaneously to SO$(3)$  for positive $m^2$, as we show explicitly in Appendix \ref{SO3EWbreak}. The TC gauge interaction is actually invariant under U$(3)\equiv$SU$(3)\times$U$(1)_{\rm
A}$, rather than just SU(3). However the U(1)$_{\rm A}$ axial symmetry is anomalous, and is therefore broken at the quantum level. This corresponds to the $\det M$ term in Eq.~\eqref{Vsu3}.
%

The components of the matrix $M\sim\eta^{T}\eta$ can be described in terms of the transformation properties of the composite states under SU$(2)_L\times$U$(1)_Y$. This notation is introduced in Table \ref{Mfields}.
\begin{table}[h]
\centering%
\begin{tabular}{|c||c||c|}
\hline 
 Field  & SU$(2)_{\text{L}}$  & U$(1)_{\text{Y}}$ \tabularnewline
\hline 
\hline 
 $\Delta\sim Q_{L}Q_{L}$  & $\square\square$  & 1 \tabularnewline
 $\sigma\sim Q_{L}U^\dagger_{R}$  & $\square$  & $-\frac{1}{2}$ \tabularnewline
 $\delta^{--}\sim U^\dagger_{R}U^\dagger_{R}$  & 1  & $-2$ \tabularnewline
\hline 
\end{tabular}\caption{\label{Mfields} Transformation properties of the component fields
of the matrix $M$ under SU$(2)_L\times $U$(1)_Y$. The complex scalars are grouped, based on their transformation properties under SU$(2)_L$, into one triplet, one doublet, and one singlet.}
\end{table}
Using this notation, the matrix $M$ is written in terms of complex scalars as
\begin{equation}
M=\left(\begin{array}{cccc}
\sqrt{2}\Delta^{++} & \Delta^{+} & \sigma^{0} \\
\Delta^{+} & \sqrt{2}\Delta^{0} & \sigma^{-} \\
\sigma^{0} & \sigma^{-} & \sqrt{2}\delta^{--}
\end{array}\right).\label{Matrix}
\end{equation}
This notation is suitable to study the vacuum, since the flavor extension sector breaks the global symmetry of the potential from SU$(3)$  down to SU$(2)\times$U$(1)$, (i.e. the EW gauge group). Next, we determine the effective Lagrangian terms of the flavor extension of 3MWT 
below scale $\Lambda_{\rm TC}$.

\subsection{Flavor Extension Sector}
\label{FeffL}

The four-fermion theory, Eq.~\eqref{4fL-2}, is given just below the SUSY breaking
scale and the techniquark condensate needs to be evolved down to the
EW scale. This is achieved by multiplying the techniquark Yukawa coupling
$y_{U}$, renormalized at the SUSY breaking scale,  with the dimensionless factor
\begin{equation}
\omega=\frac{\langle U_{L}U^\dagger_{R}\rangle_{m_{\textrm{SUSY}}}}{\langle U_{L}U^\dagger_{R}\rangle_{\Lambda_{\rm TC}}}=\left(\frac{m_{\textrm{SUSY}}}{\Lambda_{\rm TC}}\right)^{\gamma},\label{omega}
\end{equation}
 written under the assumption that the anomalous dimension $\gamma$
of the techniquark mass operator is constant.

Note that in the following we neglect the contribution of the last term in Eq.~\eqref{4fL-2} because that term respects the global SU$(3)$ symmetry, and therefore its effects should already be parametrized by the quartic couplings in the TC effective Lagrangian, Eq.~(\ref{Vsu3}). The masses of the SM fermions and the fourth family leptons arise from the terms on the first line of  Eq.~\eqref{4fL-2}: more specifically those masses are generated by the following four-fermion operator 
\begin{equation}
\eta^{T}K\eta\ ,
\end{equation}
 with 
\bea
K_{ij}&=&\frac{y_{U}c_{\theta}\omega}{m_{s}^{2}}\left[\delta_{ik}c_{\theta}\left(q_{Lu}^{\dagger k}Y_{u}^{*}u_{R}+y_{N}^{*}L_{L}^{\dagger k}N_{R}\right)-\epsilon_{ik}s_{\theta}\left(q_{Ld}^{k}Y_{d}d^\dagger_{R}+l_{L}^{k}Y_{l}e^\dagger_{R}+y_{E}L_{L}^{k} E^\dagger_{R}\right)\right]\delta_{3 j},\nonumber\\
i, j &=& 1,\ldots ,3;\quad  k=1,2;\quad \epsilon_{3 k}\equiv 0\ ,
\label{Zsp}\eea
upon condensation of the techniquarks. Under SU$(3)$ global symmetry the spurion $K$ transforms as 
$K\rightarrow {\rm u}^{*}K {\rm u}^{\dagger}$.

The four-techniquark term on the other hand is
\begin{equation}
\frac{y_{U}^{2}c_{\theta}^{2}}{m_{s}^{2}}\omega^{2}(Q_{L}U^\dagger_{R})(Q_{L}^{\dagger}U_{R})=K'_{ijkl}\eta_{i}^{\alpha}\eta_{j\alpha}\eta_{k\beta}^{\dagger}\eta_{l}^{\dagger\beta}\;\;,\;\; K'_{ijkl}=\frac{y_{U}^{2}c_{\theta}^{2}}{m_{s}^{2}}\omega^{2}\left(\delta_{ik1}+\delta_{ik2}\right)\delta_{jl3},\label{4ETC}
\end{equation}
where $\alpha$ and $\beta$ are spin indices. For this term to be
invariant under SU$(3)$, the spurion $K'$ must transform as $K'_{ijkl}\rightarrow {\rm u}_{im}{\rm u}_{jn}{\rm u}_{ko}^{*}{\rm u}_{lp}^{*}K'_{mnop}$  , with ${\rm u}\in$~SU$(3)$. To estimate the effects of renormalization, we simply assumed factorization, leading to a multiplicative factor of $\omega^{2}$.

At the lowest order in the spurions, the SU$(3)$ breaking effective Lagrangian, obtained from Eqs.~(\ref{Zsp},\ref{4ETC}) is: 
\begin{equation}
\mathcal{L}_{F}=c_1\Lambda_{\rm TC}^2\text{Tr}\left[MK\right]
+c_2\Lambda_{\rm TC}^4 K'_{ijkl}M_{ij}M_{kl}^{*}+{\rm{h.c.}}\ .\label{eeLETC}
\end{equation}
The factors of $\Lambda_{\rm TC}$ have been added to make dimensionless the coefficients $c_i$, which parametrize the uncertainty in the couplings of the effective Lagrangian in terms of those of the underlying theory. We estimate these coefficients using dimensional analysis \cite{Georgi:1992dw,Cohen:1997rt,Antola:2011at} and find
\be
c_1={\cal O}\left(\Upsilon^{-1}\right)\ ,\quad c_2={\cal O}\left(\Upsilon^{-2}\right)\ ,\quad \Upsilon\equiv \frac{\Lambda_{\rm TC}}{v_w}\ .
\ee
Finally, we note that, due to the last term in Eq.~\eqref{eeLETC},
the global SU$(3)$ symmetry 
breaking pattern of the TC potential is altered compared to the pure TC scenario presented in Appendix~\ref{SO3EWbreak}. We will study the ground state of the full potential in Section~\ref{MEig}. 
To conclude this section, we next discuss how to introduce also composite vector fields consistently. 

\subsection{Vector Sector}
\label{VeffL}

Similarly to QCD, in TC a tower of composite states is predicted to arise at low energies. 
The lightest states of the spectrum are constituted by the scalar composite states, already introduced in Subsection~\ref{SeffL},  and the vector resonances. 
A minimal set of composite vector fields transforming homogeneously under SU$(3)$ can be written, in terms of Gell-Mann matrices $\lambda^a$, as
\be\label{Vmtx}
A_\mu=A_\mu^a \frac{\lambda^a}{2}\ ,
\ee
which transform under SU(3) according to
\be
A_\mu\rightarrow {\rm u} A_\mu {\rm u}^\dagger \ , \qquad {\rm with} \qquad {\rm u}\in {\rm SU(3)} \ .
\ee
The elementary particle content of $A^\mu$ is expressed by the equivalence
\be\label{Vcont}
A^{\mu j}_i\sim\eta_i^\alpha \sigma^\mu_{\alpha\dot{\beta}} \eta^{\dagger\dot{\beta} j}-\frac{1}{3}\delta_i^j \eta_k^\alpha \sigma^\mu_{\alpha\dot{\beta}} \eta^{\dagger\dot{\beta} k}=\overline{\Psi}^j\gamma^\mu\Psi_i -\frac{1}{3}\delta_i^j\overline{\Psi}^k\gamma^\mu\Psi_k\ ,\quad
\overline{\Psi} =\left( \bar{U}_L, \bar{D}_L, \bar{U}^{\rm c}_R  \right)\ ,
\ee
where the components of $\Psi$ in SU$(3)$ space are Dirac spinors, with the superscript $^c$ on the last entry denoting the charge conjugation. The vector and axial-vector charge eigenstates and their elementary particle content are given in Appendix~\eqref{generators}.

The effective Lagrangian including composite vector fields, $A^\mu$, has already been derived in \cite{Foadi:2007ue} by applying the hidden local symmetry principle \cite{Bando:1984ej,Bando:1987br} for MWT, which features an SU$(4)$ global symmetry in the TC sector. Those results can be straightforwardly used for SU$(3)$ symmetric TC by defining the corresponding vector field and the field strength tensor:
\be\label{Cmu}
C_\mu=A_\mu-\epsilon ~ G_\mu\ ,\quad \epsilon=\frac{g_L}{g_{\rm TC}} \ ,\quad F_{\mu\nu}=\partial_\mu A_\nu-\partial_\nu A_\mu-i g_{\rm TC} \left[A_\mu,A_\nu\right]\ ,
\ee
with $G_\mu$ defined in Eq.~\eqref{Gm}. The vector field $C_\mu$ has the same transformation law as $A_\mu$:
\be
C_\mu\rightarrow {\rm u} C_\mu {\rm u}^\dagger \ , \qquad {\rm with} \qquad {\rm u}\in {\rm SU(3)} \ .
\ee
The kinetic and mass terms for the vector fields can then be written as:
\be\label{Wk}
{\cal L}_k\supset -\frac{1}{2}{\rm Tr}\left[\tilde{W}_{\mu\nu}\tilde{W}^{\mu\nu}\right]-\frac{1}{4}B_{\mu\nu}B^{\mu\nu}-\frac{1}{2}{\rm Tr}\left[F_{\mu\nu}F^{\mu\nu}\right]+m_A^2{\rm Tr}\left[C_\mu C^\mu\right]\ ,
\ee
while the scalar-vector field interaction terms up to dimension four operators read:
\be\label{LVM}
{\cal L}_{M \mbox{-} V}=g_{\rm TC}^2 r_1{\rm Tr}\left[C_\mu C^\mu M M^\dagger\right]+g_{\rm TC}^2 r_2{\rm Tr}\left[C_\mu M  C^{\mu T} M^\dagger\right]-g_{\rm TC}^2 \frac{r_3}{4}{\rm Tr}\left[C_\mu C^\mu\right] {\rm Tr}\left[ M M^\dagger\right]\ ,
\ee
with the $r_i$ constants being of ${\cal O}(1)$.
A few remarks are in order: First, higher dimensional operators are suppressed by positive powers of $\Lambda_{\rm TC}$, and therefore are subleading. Second, terms proportional to $y_U$, which explicitly break SU$(3)$ global symmetry, are small compared to those proportional to $g_{\rm TC}^2$ and therefore negligible at leading order. Third, to simplify the phenomenological analysis of 3MWT, presented in the next section, we neglected also a covariant derivative coupling term (see \cite{Foadi:2007ue} for its precise definition).\footnote{Neglecting this
term is simply a restriction on the parameter space: this term could be included in more thorough future analyses.}

\section{Phenomenology\label{pheno}}

In this section we determine the physical states arising at low energy in 3MWT, with particular focus on the light composite Higgs boson, and then study their contributions to the EW precision parameters $S$ and 
$T$.

\subsection{EW Symmetry Breaking \& Mass Eigenstates\label{MEig}}

The most general ground state  which 
breaks the EW gauge group down to the electromagnetic U$(1)_Q$
can be parametrized by the following form of the vacuum expectation value (vev) of the matrix field $M$
\begin{equation}
\langle M\rangle=\frac{1}{\sqrt{2}}\left(\begin{array}{ccc}
0 & 0 & v_{\sigma} \\
0 & \sqrt{2}v_{\Delta} & 0 \\
v_{\sigma} & 0 & 0 
\end{array}\right)\ .\label{Mvev}
\end{equation}
Minimizing the scalar potential given in Eqs.~(\ref{Vsu3},\ref{eeLETC})
leads to the following relations between the parameters: 
\be
m^{2}=\lambda\left(v_{\sigma}^{2}+v_{\Delta}^{2}\right)+4\lambda'v_{\Delta}^{2}-2 \lambda'' v_{\sigma}^{2}, \qquad\tilde{m}^{2}=2\left(\lambda'+\lambda''\right)\left(v_{\sigma}^{2}-2v_{\Delta}^{2}\right),\label{vevc}
\ee
where we have denoted 
\begin{equation}
\tilde{m}^{2}=c_2y_{U}^{2}c_{\theta}^{2}\frac{\Lambda_{\rm TC}^4\omega^{2}}{m_{s}^{2}}\ ,\quad \lambda''\equiv-\frac{m'}{v_{\Delta}}.\label{eq:mFE}
\end{equation}
A sufficient condition for the potential to be bounded from below is $\lambda,\lambda'>0$. 
In the limit $v_\sigma=\sqrt{2}v_\Delta$ we obtain $\tilde{m}=0$, consistently with the pure TC result in Appendix~\ref{SO3EWbreak}. We notice that the ground state changes because of the four-technifermion interaction in Eq.~\eqref{4ETC}. An analogous change of ground state occurs in ETC theories where some of the chiral symmetries of the pure TC theory are broken by extended gauge interactions.
In our model setup the effective four fermion interactions at low energy arise from attractive 
Yukawa couplings which is different from the usual ETC scenario where the underlying gauge intractions
can be either repulsive or attractive.

The mass spectrum of 3MWT, corresponding to the vev in Eq.~\eqref{Mvev}, includes two neutral scalars, $h^0$ and $H^0$, as well as one neutral pseudoscalar,  $\Pi^0$, one charged and two doubly charged scalars, $H^\pm $, $h^{\pm\pm}$, and $H^{\pm\pm}$, respectively, with the corresponding squared mass matrices given in Appendix~\ref{sMMx}. We introduce for later use the mixing angle $\varphi$ defined by
\be
\begin{pmatrix} h^0 \\ H^0 \end{pmatrix} = \frac{1}{\sqrt{2}}\begin{pmatrix} c_\varphi & -s_\varphi \\ s_\varphi & c_\varphi \end{pmatrix} \begin{pmatrix} \Re\left(\sigma^0\right) \\ \Re\left(\Delta^0\right) \end{pmatrix} ~. \
\label{ScalarMix}
\ee

Assuming that only the third generation SM fermions have non-negligible Yukawa couplings, it follows from Eq.~$\eqref{eeLETC}$ that the masses of the upper component $u$ and the lower component
$d$ of a generic SM fermion EW doublet are given, respectively, by 
\begin{equation}\label{mu}
m_{u}=c_1\frac{c_{\theta}^{2}y_{U}y_{u}\omega \Lambda_{\rm TC}^2}{m_{s}^{2}}\frac{v_{\sigma}}{\sqrt{2}},\qquad\qquad m_{d}=\frac{y_{d}}{y_{u}}t_{\theta}m_{u}\ .
\end{equation}
Using the previous equation, the fact that both and $E$ and $N$ have to be heavier than about 100~GeV, and requiring the Yukawa couplings to be perturbative, we deduce that $\theta$ cannot be close to either 0 or $\pi/2$. To simplify the study of 3MWT from here on we take $m_s=m_{\rm SUSY}$, which from Eq.~\eqref{eq:mss} is equivalent to imposing:
\be
m_{\rm SUSY}=m_s \quad\Rightarrow \quad \frac{b}{\mu^2}=\frac{t_{\theta}+t^{-1}_\theta}{2}\ .
\ee
From the fact that neither $t_{\theta}$ nor its inverse are large, it follows that our choice does not introduce any large hierarchy between $b$ and $\mu^2$, which can be taken both of ${\cal O}(m^2_s)$. 
This implies in particular that the higgsinos are heavy and decoupled from low energy phenomenology.

Finally, the composite axial-vector and vector resonances mix with the SM gauge bosons, while the doubly charged baryon technivector does not. The resulting physical massive vector states are $Z_\mu$, $Z'_\mu$, $Z''_\mu$, $W^\pm_\mu$, $W'^\pm_\mu$, $W''^\pm_\mu$, $\Omega^{\pm\pm}_\mu$, with the corresponding squared mass matrices given in Appendix~\ref{sMMx}.
The masses of $W^\pm$ and $Z$ in the limit of negligible mixing ($\epsilon=0$) read:
\be
m_W^2\cong\frac{1}{4}g_L^2 v^2_w\\ ,\quad m_Z^2\cong\frac{1}{4}(g_L^2+g_Y^2) (1+t_\rho^2) v^2_w \ ,
\ee
where  $t_\rho=\sqrt{2} v_\Delta/v_\sigma$, and the EW scale is given by
\be\label{vsD}
v_w^2=(\sqrt{2}G_F)^{-1}=(246\ {\rm GeV})^2=v_{\sigma}^2+2v_{\Delta}^2\ .
\ee

\subsection{Higgs Mass}

The light Higgs deserves some further discussion. In \cite{Foadi:2012bb} it has been shown that $m^2_{h^0}$ receives a large negative corrections at one loop from the top quark and the $W^\pm$ and 
$Z$ bosons. Because of these corrections the tree level 
Higgs mass can be significantly larger than 125 GeV: Adapting the formulas given in \cite{Foadi:2012bb} to the 3MWT case, while neglecting in first approximation any mixing between the two neutral scalar states, we can write
\be
\left(m^2_{h^0}\right)_{\rm{tree}}
\simeq m^2_{h^0}+\frac{8}{3}\kappa^2\left[ 2 a_f^2 \left( 3 m_t^2+m_E^2+m_N^2  \right) - 3 a^2_\pi \left( m_W^2 + \frac{m_Z^2}{2}  \right) \right]\ ,
\ee
where $(m^2_{h^0})_{\rm{tree}}$ is the scalar mass due to pure strong dynamics before coupling with the EW gauge currents and SM matter. Furthermore, $\kappa$ is an ${\cal O}(1)$ renormalization coefficient, while $a_f$ and $a_\pi$ are rescaling coefficients of the SM Higgs linear coupling to fermions and quadratic coupling to gauge bosons, respectively. In writing the formula above we set the dimension of the TC representation to 3 and the number of technidoublets to 0.75, as it is the case for 3MWT. To illustrate the idea we evaluate the formula above by assuming couplings and renormalization coefficient to be SM like  ($a_F=a_\pi=\kappa=1$), and $m_E=m_N=m_t$. With these choices we find a light Higgs mass equal to 740 GeV. This value is not far from the naive TC estimate for 3MWT, obtained by scaling up the mass of the $f_0(500)$ QCD resonance \cite{Beringer:1900zz} ($m_{f_0}$=400-550 GeV):
\be
\left(m^2_{h^0}\right)_{\rm{naive}}\simeq \frac{4}{3}\frac{v_w^2}{f_\pi^2}m_{f_0}^2 = \mbox{\rm 1200-1700 GeV} \ .
\ee
The required suppression of the tree level Higgs mass, around 50\%, might realistically come from near-conformal dynamics. This should be indeed the case for 3MWT, which has 1.5 adjoint Dirac flavors and is outside but close to the conformal window  \cite{Dietrich:2006cm}.
Lattice results moreover suggest that for 2 Dirac flavors the lower bound of the conformal window extends below the analytic estimates based on two loop perturbative beta function and ladder approximation of the Schwinger-Dyson equation for the fermion propagator. Therefore the suppression of the tree level Higgs mass in our theory could be more pronounced than what perturbative estimates might suggest.

Taking into account one loop corrections to the mixing of the two neutral scalar states would require us to introduce the one loop effective potential: this is beyond the task of our study, which is to perform an initial investigation of the viability of 3MWT. For this reason we simply take the tree level light Higgs mass to be 125 GeV, by assuming that this value incorporates not only the strong dynamics contribution but receives also non-leading corrections from the global SU$(3)$ breaking sector of 3MWT. 

\subsection{Oblique Corrections\label{Obl}}

The precision EW parameters \cite{Peskin:1990zt,Peskin:1991sw} can be calculated directly from the vector-boson sector of the effective Lagrangian, Eqs.~(\ref{Wk},\ref{LVM}), by integrating out the heavy charged and neutral states and then using the formulas provided by \cite{Barbieri:2004qk,Chivukula:2004pk}.\footnote{For this task we adapted the code provided by the authors of \cite{Foadi:2007ue} to 3MWT.} At tree level and linear order in the mixing parameter $\epsilon$ we find:
\be
S_{tree}=0\ ,\quad \alpha_e T_{tree} = -\frac{2v_{\Delta}^2}{v_w^2}\ .\label{Ttree}
\ee
For the $T$-parameter to be consistent with the experiments, the vev component $v_{\Delta}$ clearly has to be small relatively to the EW scale. We note that the $S$ parameter is zero up to corrections of order $\epsilon^4$ while the $T$ parameter obtains further contributions of order $\epsilon^2$ which we neglect since $\epsilon\ll 1$.

The intrinsic TC contribution is usually calculated from the one loop perturbative diagrams of the technifermions, which are assumed to be massive because of dynamical 
symmetry breaking. The dynamical mass divided by $m_Z$ is usually taken infinite,  as this gives a meaningful result with no unknown parameters. These contributions are denoted by $S_{naive}$ and $T_{naive}$. For our underlying technicolor theory, Eq.~\eqref{3MWTQ}, the dynamical masses 
are such that while the up-techniquark, $U$, gains only a Dirac mass, $m_U$, the down-techniquark, 
$D_L$, acquires also a Majorana mass, $m_L$. Oblique corrections for this general case have been calculated in \cite{Antipin:2009ks} in terms of integral functions, which we used to derive the explicit formulas given in Appendix~\ref{STgen}. In the pure 3MWT limit the Dirac and Majorana masses 
are equal to each other\footnote{This follows from the invariance of the mass terms implied by Eq. (\ref{Matrix}) under SO(3).}, $m_L=m_U$, and we find
\be\label{STMaj}
S_{naive}=\frac{3}{4\pi}\;\;,\;\;T_{naive}=\frac{m_U^2\log[\frac{m_U}{\Lambda_{NP}}]}{4m_Z^2s^2_{\xi}c^2_{\xi} \pi}\ .
\ee
The dependence of $T$ on the renormalization scale $\Lambda_{NP}$ 
is due to the Majorana mass, and it should be matched onto a renormalizable term in the underlying theory.
For the phenomenological purposes of our current analysis, we assume that the renormalization scale 
$\Lambda_{NP}$ is close to $m_U$ making the one-loop contribution $T_{naive}$ negligible when compared to the tree-level contribution, $T_{tree}$.\footnote{A generic renormalization scale still allows viability since $T_{naive}$ can be canceled by $T_{tree}$, though this requires fine tuning of $v_\Delta$. 
} With this assumption we use the following values of the naive parameters:
\be
S_{naive}=\frac{3}{4\pi} \;\;,\;\;T_{naive}=0\label{Snaive}\ .
\ee

Another independent contribution to $S$ and $T$ in our model comes from the fourth family leptons $N$ and $E$. We denote these contributions by $S_{N,E}$ and $T_{N,E}$. These have been already evaluated in \cite{He:2001tp}: for $m_{E}>m_N$, the $S_{N,E}$ contribution is actually negative and can offset the positive $S_{naive}$. We can therefore summarize the non-negligible contributions to the $S$ and $T$ parameters, with the assumptions described above, as:
\be
S=S_{naive}+S_{N,E} \;\;,\;\;T=T_{tree}+T_{N,E},
\ee
where $T_{tree}$, $S_{naive}$, and the Higgs contributions are given in Eqs.~(\ref{Ttree},\ref{Snaive}).

\section{Experimental Validation\label{exp}}

In this section we study the viability of 3MWT by performing a goodness of fit analysis based on the recent LHC and Tevatron data on Higgs physics, as well as the experimental values of the $S$ and $T$ parameters. In Subsection~\ref{ccoeff} we derive the linear couplings of the light Higgs and relate these to the corresponding SM coupling strengths. In Subsection~\ref{PSS} we implement the direct search lower bounds on the new physics mass spectrum by performing a numerical scan over the parameter space and collecting data points that satisfy these constraints and the EW precisions tests. Finally, in Subsection~\ref{GoF} we define the observables included in the goodness of fit analysis and then present the statistical results on the viability of 3MWT.

\subsection{Coupling Coefficients}\label{ccoeff}

In our model the linear Higgs coupling to charged vector bosons can be written in compact form as
\be\label{Hint}
{\cal L}\supset \frac{2 m_A^2}{v_w}\bar{W}^\dagger_\mu\Xi\bar{W}^\mu h^0\ ,\quad \bar{W}^ \dagger_\mu=\left(\tilde{W}^-_\mu,V_\mu^-,A_\mu^-,\Omega_\mu^{--}\right)\, ,
\ee
with the non-zero terms of the matrix $\Xi$ given by
\bea
\Xi_{1,1}&=&\left(x^2+\epsilon ^2 z_1\right) \left(c_{\varphi } c_{\rho }-\sqrt{2} s_{\varphi } s_{\rho }\right)+\frac{\epsilon ^2 z_3}{\sqrt{2}} \left(s_{\varphi }
   s_{\rho }-\sqrt{2} c_{\varphi } c_{\rho }\right)\ ,\nonumber\\ 
   \Xi_{2,2}&=&z_1 \left(c_{\varphi } c_{\rho }-\sqrt{2} s_{\varphi } s_{\rho }\right)+z_2 \left(c_{\varphi }
   s_{\rho }-\sqrt{2} c_{\rho } s_{\varphi }\right)+\frac{z_3}{\sqrt{2}} \left(s_{\varphi } s_{\rho }-\sqrt{2} c_{\varphi } c_{\rho }\right)\ ,\nonumber\\ 
   \Xi_{3,3}&=&z_1 \left(c_{\varphi } c_{\rho }-\sqrt{2} s_{\varphi } s_{\rho }\right)-z_2 \left(c_{\varphi } s_{\rho }-\sqrt{2}
   c_{\rho } s_{\varphi }\right)+\frac{z_3}{\sqrt{2}} \left(s_{\varphi } s_{\rho }-\sqrt{2} c_{\varphi } c_{\rho }\right)\ , \nonumber\\
   \Xi_{1,2}&=&-\frac{\epsilon}{2\sqrt{2}}    \Xi_{2,2}\ ,\quad
\Xi_{1,3}=-\frac{\epsilon}{2\sqrt{2}} \Xi_{3,3}\ ,\nonumber\\
  \Xi_{4,4}&=&2 c_{\varphi } c_{\rho } \left(z_1+z_2\right)+\frac{z_3}{\sqrt{2}} \left(s_{\varphi } s_{\rho }-\sqrt{2} c_{\varphi } c_{\rho
   }\right)\, . 
 \eea
 Here $\varphi$ is the mixing angle of the neutral scalars, Eq.~\eqref{ScalarMix}, and 
 \be
\epsilon=\frac{g_L}{g_{\rm{\rm TC}}},\quad x=\frac{g_L v_w}{2 m_A}\ ,\quad t_\rho=\frac{\sqrt{2} v_\Delta}{v_\sigma}\ ,\quad z_i= \left(\frac{g_{\rm TC} v_w}{2 m_A}\right)^2 r_i\ ,\ i=1,2,3\ .
\ee

To simplify the analysis we select the slice of parameter space where the axial vector coupling to the light Higgs is zero ($\Xi_{1,3}=\Xi_{3,3}=0$). Moreover, we require the mixing mass term of the axial-vector to arise only from mixing ($({\cal M}^2_W)_{1,3}=-\epsilon m_A^2/\sqrt{2} $), which is taken to be small. Together these conditions are satisfied by imposing
\be\label{PSslice}
z_1=\frac{1}{4} \left(3+c^{-1}_{2 \rho }\right) z_3\ ,\quad z_2=\frac{1}{4} t_{2 \rho } z_3\ .
\ee
If we apply the above substitution together with
\be\label{simP}
z_3\rightarrow \frac{4 s^2}{s_{2 \rho}t_{2 \rho}}
\ee
to the charged vector boson squared mass matrix in Eq.~\eqref{sqMVc}, we reproduce the corresponding result quoted in \cite{Alanne:2013dra}:
\be\label{sMMV}
{\cal M}_W^2=\left(
\begin{array}{ccc}
 m_{\tilde{W}}^2 & -\frac{\epsilon  m_V^2}{\sqrt{2}} & -\frac{\epsilon  m_A^2}{\sqrt{2}} \\
 -\frac{\epsilon  m_V^2}{\sqrt{2}} & m_V^2 & 0 \\
 -\frac{\epsilon  m_A^2}{\sqrt{2}} & 0 & m_A^2
\end{array}
\right)\ ,
\ee
with
\be
m_{\tilde{W}}=\left[x^2+\left(1+s^2\right) \epsilon ^2\right] m_A^2\ ,  \quad  m_V^2=\left(1+2 s^2\right) m_A^2\  .
\label{svar2}\ee
The mass eigenvalues can be expanded in $x$
and $\epsilon$,
which in TC are both expected to be small:
\bea\label{sMV}
m_W^2 &\cong& m_A^2 x^2\left[1-\epsilon ^2\right]\ ,\quad m^2_{W''}\cong m_A^2 \left[1+\frac{1}{2} \left(1+x^2\right) \epsilon ^2-\frac{1}{8} \left(2+\frac{1}{s^2}\right) \epsilon ^4\right]\ ,\nonumber\\
m_{W'}^2 &\cong& m_A^2 \left[1+2 s^2+\frac{1}{2} \left(1+2 s^2+x^2\right) \epsilon ^2+\frac{1}{8} \left(2+\frac{1}{s^2}\right) \epsilon ^4\right]\ ,
\eea
where we kept contributions up to ${\cal O}(x^n\epsilon^{4-n})$, with $n=0,\ldots,4$. 

Generally the linear couplings of the light Higgs can be conveniently expressed in terms of coupling coefficients defined by
\bea\label{HlinL}
{\cal{L}}_{\textrm{eff}} &=& \sum_i a_{W_i}\frac{2m^2_{W_i}}{v_w}hW^{+}_{i \mu} W^{-\mu}_i+\sum_j a_{Z_j}\frac{m^2_{Z_j}}{v_w}h Z_{j \mu} Z^{\mu}_j
\nonumber \\
&-&a_f\sum_{\psi=t,b,\tau,N,E}\frac{m_\psi}{v_w}h\bar{\psi}\psi-\sum_k a_{S_k}\frac{2m_{S_k}^2}{v_w}hS_k^+ S_k^-\ , 
\label{efflagr}
\eea
where the indices $i$ and $k$ run over all the charged scalars and vector bosons (including the states with double charge).
By substituting the charged vector mass eigenstates obtained from Eq.~\eqref{sMMV} in Eq.~\eqref{Hint}, we find the linear Higgs couplings to the charged vector bosons. By normalizing these couplings according to Eq.~\eqref{HlinL}, with the masses given by Eqs.~(\ref{sMV},\ref{smO}), we determine
\bea\label{aWp}
a_{W}&=&\left(c_{\varphi } c_{\rho }-\sqrt{2} s_{\varphi } s_{\rho }\right) \left[1-\frac{x^2 \epsilon ^2}{2}  \left(1+\frac{1}{1+2 s^2}\right)\right]+\frac{x^2
   \epsilon ^2 s^2}{2 \left(1+2 s^2\right)^2} \left(\frac{c_{\varphi }}{c_{\rho }}-\sqrt{2}\frac{ s_{\varphi }}{s_{\rho }}\right)\ ,\nonumber\\
a_{W}&+&a_{W'}+a_{W''}=c_{\varphi } c_{\rho }-\sqrt{2} s_{\varphi } s_{\rho }+\frac{s^2}{1+2 s^2}  \left(\frac{c_{\varphi }}{c_{\rho }}-\sqrt{2}\frac{ s_{\varphi }}{s_{\rho }}\right) \ ,\nonumber\\
a_{\Omega}&=&\frac{s^2}{s_{\rho }}\frac{2 c_{\varphi } \left(1+t^{-1}_{\rho }\right)+\sqrt{2} c_{2 \rho } c_{\rho }^{-2} s_{\varphi }}{2-4 s^2+s^2 \left(2
   \left(t^{-1}_{\rho }+s_{\rho }^{-2}\right)+c_{\rho }^{-2}\right)}\ .
\eea
The corresponding result for fermions reads simply
\be
a_f=\frac{c_\varphi}{c_\rho}\ .
\ee
Among the neutral vector resonances only $Z$ is relevant for the LHC observables we include in this study. As we discuss in the next section, the numerical values of $a_Z$ that we find are very close to $a_W$, and for all practical purposes they can therefore be taken equal to each other.\footnote{The analytic expressions for $a_{Z_j}$ and $a_{S_k}$ are lengthy and we do not reproduce them here.}

\subsection{Parameter Space Scan}\label{PSS}

We perform a scan of the parameter space to find data points that satisfy experimental constraints on the new physics mass spectrum and on the $S$ and $T$ parameters. We require perturbativity of all but the $g_{\rm TC}$ coupling
and stability of the potential at large values of the scalar fields. The unknown parameters derived from strong dynamics are estimated using dimensional analysis \cite{Georgi:1992dw,Cohen:1997rt,Antola:2011at}. Finally, we fix the anomalous dimension to 
$\gamma=1.5$, and require the supersymmetry breaking scale to be larger than 5 TeV. Putting all this together, the free parameters in our scan acquire values in the domain defined by the following relations:
\bea
&& 210~{\rm GeV} \leq |v_\sigma| \leq 246~{\rm GeV}\,,\ m_A=1~{\rm TeV}\,,\ \gamma=1.5\,,\ \pi\leq \Upsilon\leq 4\pi\,,\ 0.5\leq c_1\Upsilon^{-1}\leq 5\,,\nonumber\\
&& 0.1 \leq \lambda \leq (2\pi)^2\,,\ 0.1\leq y_t,y_N,y_E,y_U\leq 2\pi\,,\ 0.1\leq \lambda''\leq 200\,,\ 0\leq z_3 \leq 1\,,\ |\epsilon|\leq2 x\,.
\eea
The remaining model parameters, $\lambda',\theta,x,v_\Delta$ are determined in terms of the ones above by using, respectively, Eqs.~(\ref{Mhn},\ref{mu},\ref{vsD},\ref{sMV}) together with the observed masses of the Higgs boson, top quark, $W$ boson, and the EW vev.
The last relation above ensures that the physical $W$ is mostly made of the SU$(2)_L$ gauge field. We 
 scanned over the parameter space defined above and collected 1000 data points, each satisfying the following constraints:
\be
m_{\rm SUSY}>5~{\rm TeV}\,,\ m_{H^0}>600~{\rm GeV}\,,\ m_{\Pi^0},m_{H^\pm},m_{h^{\pm\pm}},m_{H^{\pm\pm}},m_E,m_N\geq 100~{\rm GeV}\, ,
\ee
as well as the experimental limits on $S$ and $T$ \cite{Beringer:1900zz}. We also checked that the collected data points satisfy the ATLAS lower limit on the mass of a sequential $W'$ boson \cite{ATLAS:2012loa}, equal to 2.55~TeV, after an appropriate rescaling of the limit which takes into account the non-SM value of the $W'$ coupling to fermions \cite{Alanne:2013dra}.

\subsection{Goodness of Fit}\label{GoF}
We performed a goodness of fit analysis by using the observed Higgs decay rates to $\gamma\gamma\,,ZZ,\,WW,\,\tau\tau,\,bb,\,\gamma\gamma JJ$ at ATLAS \cite{ATLAS-CONF-2012-160,ATLAS-CONF-2012-161,ATLAS-CONF-2013-030,ATLAS-CONF-2013-012,ATLAS-CONF-2013-013} and CMS \cite{CMS-PAS-HIG-13-002,CMS-PAS-HIG-13-001,CMS-PAS-HIG-13-003,CMS-PAS-HIG-13-004}, and to $\gamma\gamma,\,WW$, and $bb$ at Tevatron \cite{Aaltonen:2013kxa}, as well as the $S$ and $T$ experimental values \cite{Beringer:1900zz}, for a total of 19 observables. The LHC and Tevatron results are expressed in terms of the signal strengths, defined as
\be
\hat{\mu}_{ij}=\frac{\sigma_{\textrm{tot}}{\textrm{Br}}_{ij}}{\sigma_{\textrm{tot}}^{\textrm{SM}} \textrm{Br} ^{\textrm{SM}}_{ij}}\ ,\quad \sigma_{\rm tot}=\sum_{\Phi'=h,qqh,\ldots}\epsilon_{\Phi'}\sigma_{\Phi\rightarrow \Phi'}\ ,\quad \Phi=pp,p\bar{p} \ ,
\label{LHCb}\ee
where $\epsilon_{\Phi'}$ is the efficiency associated with the given final state $\Phi'$ in an exclusive search, while for inclusive searches one simply has $\sigma_{\rm tot}=\sigma_{pp\rightarrow h^0(X)}$, the $h^0$ production total cross section.

The combined signal strengths from ATLAS, CMS\footnote{We use the mass cut based result for CMS result on the Higgs to diphoton decay.}, and Tevatron are given in Table~\ref{datatable},
\begin{table}[htb]
\begin{tabular}{|c||c|c|c|}
\hline
$ij$ & ATLAS & CMS & Tevatron \\
\hline
$ZZ$ & $\,1.50\pm 0.40\,$ & $\,0.91\pm 0.27\,$ &  \\
 $\gamma\gamma$ & $1.65\pm 0.32$ & $1.11\pm 0.31$ & $6.20\pm 3.30$ \\
$WW$ & $1.01\pm 0.31$ & $0.76\pm 0.21$ & $0.89\pm 0.89$ \\ 
$\tau\tau$ & $0.70 \pm 0.70$ & $1.10\pm 0.40$ & \\
$bb$ & $-0.40\pm 1.10$ & $1.30\pm 0.70$ & $1.54\pm 0.77$\\
\hline
\end{tabular}
\caption{Combined signal strengths from LHC and Tevatron experiments.}
\label{datatable}
\end{table}
while the signal strengths and efficiencies\footnote{We chose to include only the loose categories from the ATLAS and CMS dataset at 8 TeV.} for dijet associated $\gamma\gamma$ production at ATLAS and CMS are listed in Table~\ref{ggjjdata}.
\begin{table}[htb]
\begin{tabular}{|c||c|c|c|c|}
\hline
 & ATLAS 7TeV & ATLAS 8TeV & CMS 7TeV & CMS 8TeV \\
\hline
$\gamma\gamma J J$ & $\,2.7\pm 1.9\, $ & $2.8\pm 1.6$ &  $2.9\pm 1.9$ &  $0.3\pm 1.3$ \\
\hline
 $pp\rightarrow h$ & $22.5\%$ & $45.0\%$ & $26.8\%$  & $46.8\%$ \\
$pp\rightarrow qqh$ & $76.7\%$ & $54.1\%$ & $72.5\%$  & $51.1\%$ \\ 
$pp\rightarrow t\bar{t}h $ & $0.6\%$ & $0.8\%$ & $0.6\%$ & $1.7\%$ \\
$pp\rightarrow Vh $ & $0.1\%$ & $0.1\%$ & $0\%$ & $0.5\%$ \\
\hline
\end{tabular}
\caption{Signal strengths and efficiencies for Higgs decay to $\gamma\gamma$ associated to a dijet at LHC.}
\label{ggjjdata}
\end{table}

Finally, the observed values for the $S$ and $T$ parameters read \cite{Beringer:1900zz}:
\be
S=0.04\pm 0.09\ ,\ T=0.07\pm 0.08\ ,\ r(S,T)=88\%\ ,
\ee
with the last quantity defining the correlation of the two parameters.

\begin{figure}[htb]
\includegraphics[width=0.48\textwidth]{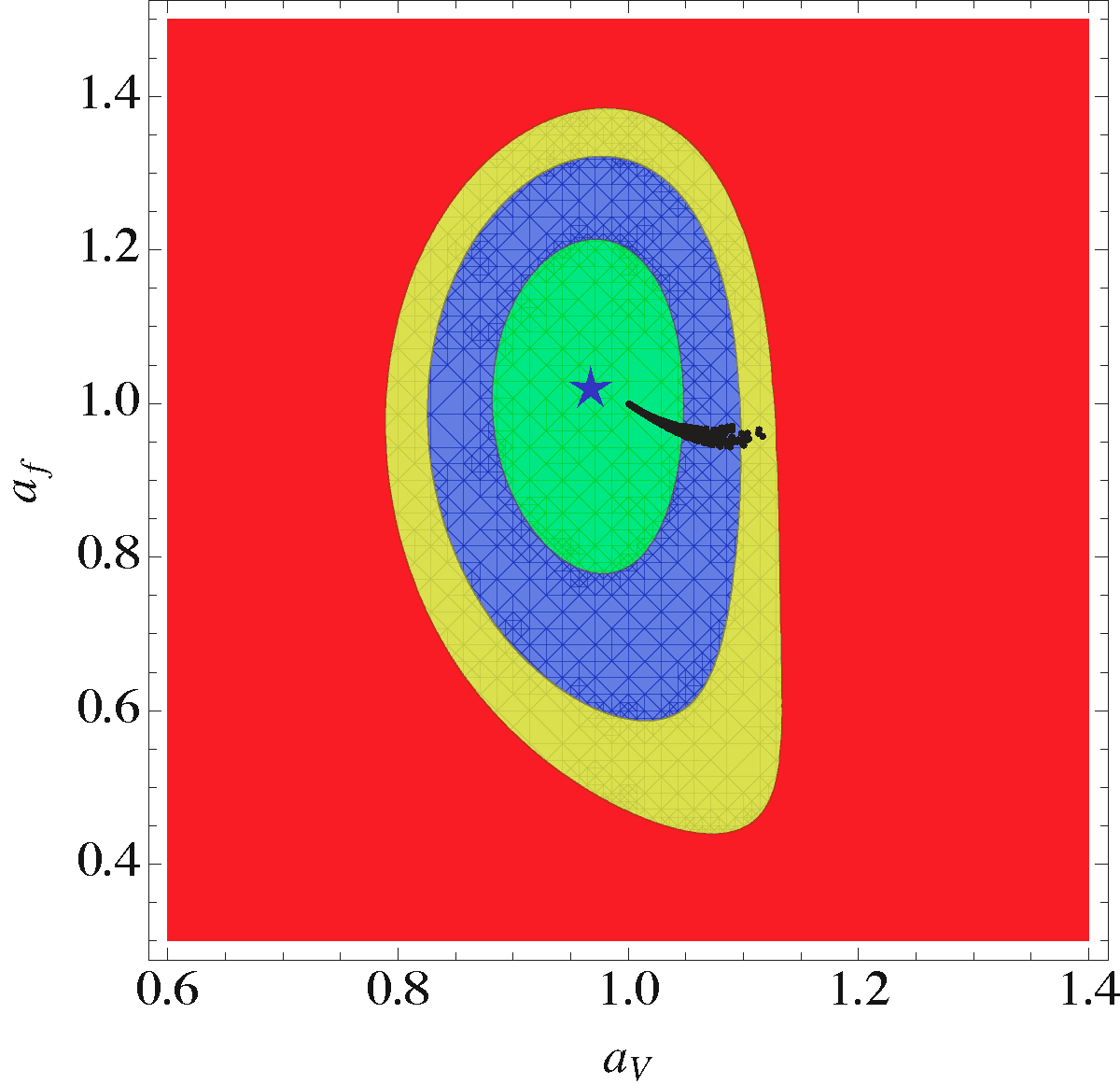}\hspace{0.45cm}
\includegraphics[width=0.48\textwidth]{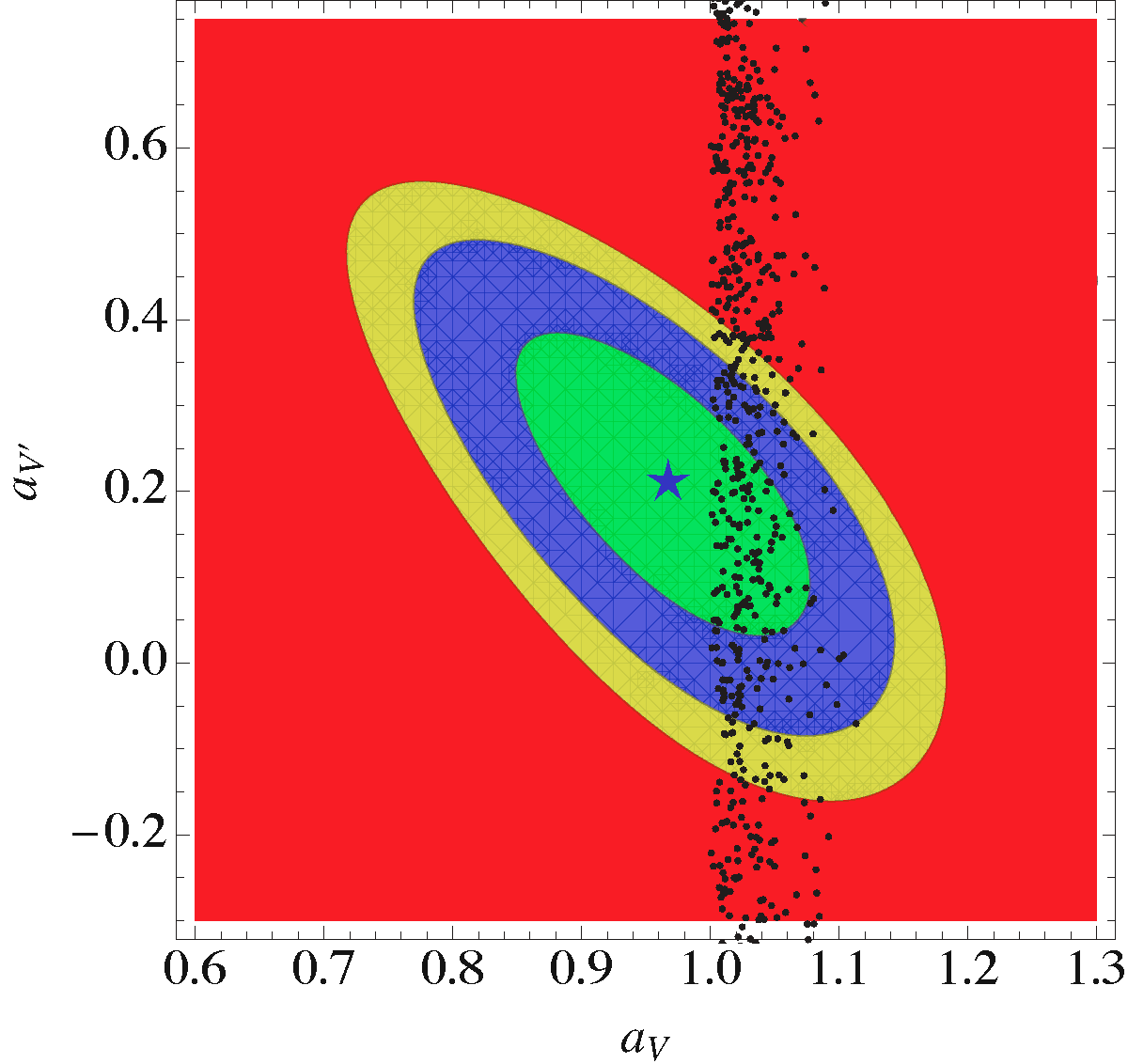}\hspace{0.45cm}
\caption{Viable data points in the $(a_V,a_f)$ (left panel) and $(a_V,a_{V'})$ (right panel) planes, together with the 68\% (green), 90\% (blue), and 95\% (yellow) CL region. The blue star in each plot marks the optimal coupling coefficients on the respective planes.}
\label{aVVpf}
\end{figure}

\begin{figure}[htb]
\includegraphics[width=0.55\textwidth]{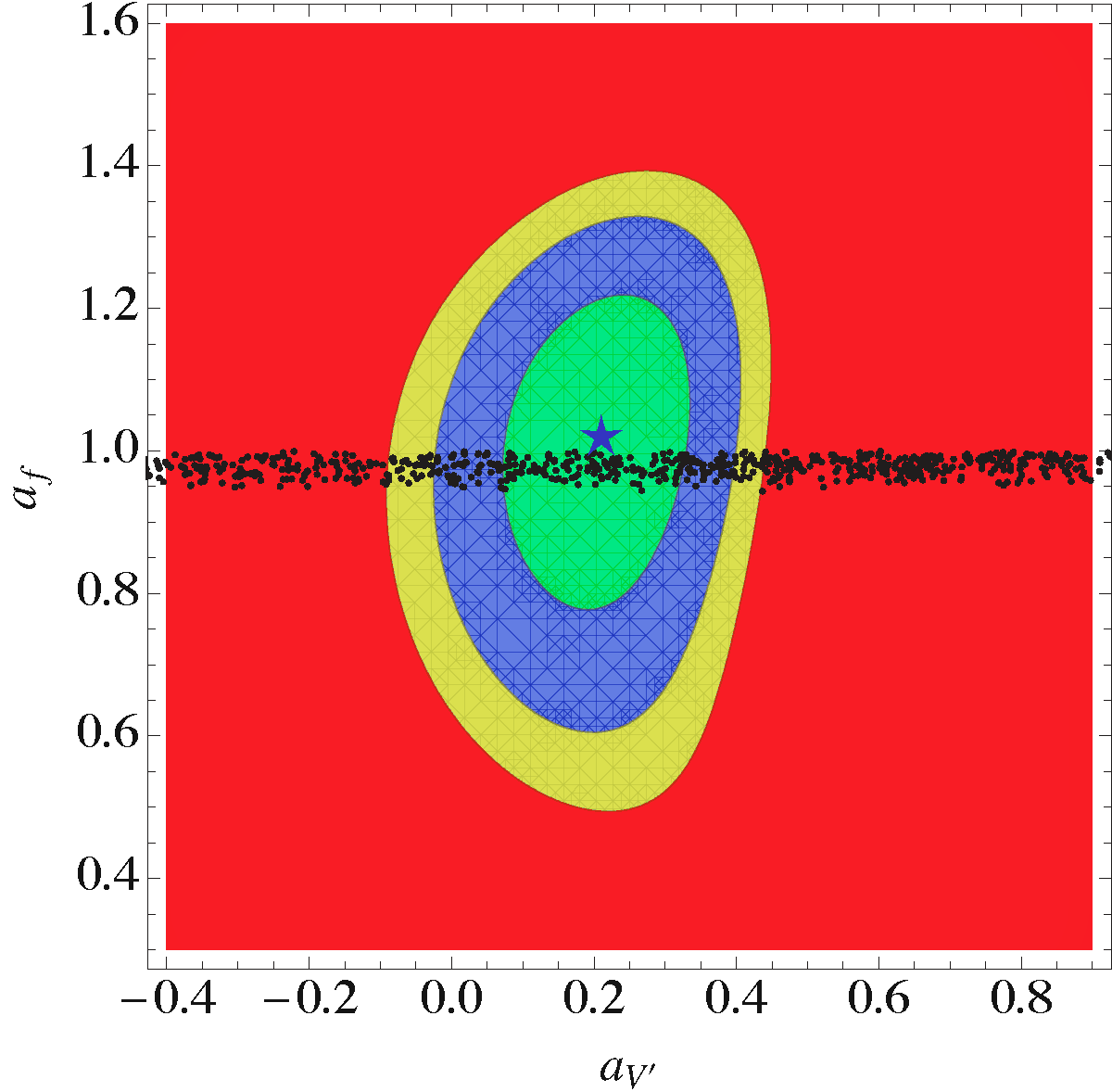}\hspace{0.5cm}
\caption{Viable data points in the $(a_{V'},a_f)$ plane passing through the point with optimal coupling coefficients in the $(a_{V'},a_f,a_V)$ space, marked by a blue star, together with the 68\% (green), 90\% (blue), and 95\% (yellow) CL region.}
\label{aVpf}
\end{figure}

For a detailed description of the present fit we refer the reader to \cite{Alanne:2013dra}, where the same statistical analysis has been performed for a different model. 
Given that no new physics has been detected, only the contributions of new charged particles at one loop to $\Gamma_{h\rightarrow \gamma\gamma}$ become relevant when comparing the 3MWT predictions to the data in Tables~\ref{datatable}, \ref{ggjjdata}. More explicitly one has \cite{Gunion:1989we}
\be\label{hgamgam}
\Gamma_{h\rightarrow \gamma\gamma}= \frac{\alpha_e^2 m_{h}^3}{256 \pi^3 v_w^2}\left| \sum_i  N_i e^2_i F_{i} \right|^2\ ,
\ee
with $i$ summed over all the charged particles, $N_i$ is the number of colors, $e_i$ the charge in electron units, and $F_i$ a function of the mass $m_i$ and the coupling coefficient defined in 
\cite{Alanne:2013dra}. In the limit of new particles being much heavier than the light Higgs, one finds
\be
F_{W_i}=7 a_{W_i}\ ,\quad F_E=F_N=-a_f\frac{4}{3}\ ,\quad F_{S_i}=-a_{S_i}\frac{1}{3}\ ,
\ee
with the coupling coefficients defined by Eq.~\eqref{HlinL}. We can therefore mimic the contribution of the charged non-SM particles in 3MWT to the observables in Tables~\ref{datatable}, \ref{ggjjdata} by including only the new contribution of a heavy single charged vector boson with coupling coefficient $a_{V'}$ determined by
\be\label{aVp}
a_{V'}\equiv \frac{1}{7}\left(F_{W'}+F_{W''}+4 F_{\Omega}\right)-\frac{a_S}{21}\ ,\quad a_S \equiv -3 \left( 16 F_E + 4 F_N + F_{H^\pm}+4 F_{h^{\pm\pm}}+4 F_{H^{\pm\pm}}\right)\ ,
\ee
where the factors of 4 account for the double charge of the corresponding states. Moreover to simplify the analysis consistently with \cite{Alanne:2013dra} we redefine
\be\label{aV}
a_Z\approx a_W\equiv a_V\ ,
\ee
where the numerical deviations from the first approximate equality above are negligible for the collected data points compared to the uncertainties on the coupling coefficients, which we present at the end of this section. At each collected data point we determine the numerical values of $a_f,a_V$, and $a_{V'}$ by Eqs.~(\ref{aWp},\ref{aVp},\ref{aV}), while we calculate numerically the coupling coefficients of the charged scalars. In Fig.~\ref{aVVpf} we plot the viable data points on the $(a_V,a_f)$ (left panel) and $(a_V,a_{V'})$ (right panel) planes together with the 68\% (green), 90\% (blue), and 95\% (yellow) confidence level (CL) regions. In both plots, the third parameter is fixed to the optimal value marked with a blue star.
It is clear from Fig.~\ref{aVVpf}, left panel, that the $W$ and $Z$ couplings are enhanced, compared to their SM values, while the SM fermion couplings are suppressed. This result for the 3MWT couplings enhances the Higgs decay to diphotons. However, the contribution of the new charged fermions and scalars, as can be seen from Eq.~\eqref{aVp}, is large and interferes destructively with the $W$ contribution to the same process. The probability associated with the data point minimizing $\chi^2$ in the $(a_f,a_V,a_{S})$ space, obtained from a 3MWT particle spectrum lacking composite vector resonances at low energy would be ruled out entirely:
\bea\label{minchi2S}
a_V&=&1.00\ ,\quad a_f=1.00\ ,\quad a_{S}=20.5\ ,\ \quad S=0.04\ ,\ T=0.07\ ;\nonumber\\ 
\chi^2_{\min}/\textrm{d.o.f.}&=&3.42\ ,\quad P\left(\chi^2>\chi_{\min}^2\right)=0.0004\,\%\ ,\quad \textrm{d.o.f.}=16\ .
\eea
In calculating $\chi^2_{\min}/\textrm{d.o.f.}$ in the the above equations we assumed the model to allow for 3 free paramters $(a_f,S,T)$, since $a_V$ is strongly correlated with $a_f$ near $\chi^2_{min}$ and $a_S$ is basically constant. Fortunately the goodness of the fit changes dramatically once we include composite vector resonances in the 3MWT low energy spectrum, as it can be guessed from Fig.~\ref{aVVpf}, right panel, and from Fig.~\ref{aVpf}.
 The contribution of the new charged vector bosons, and especially that of the vector baryon $\Omega$, to the Higgs decay into diphoton is large, and offsets entirely the negative contribution of $E$, $N$, and charged scalars in Eq.~\eqref{aVp}. Among the 1000 viable data points, the one producing the minimum value for $\chi^2$ gives:
\bea\label{minchi2Vp}
a_V&=&1.01\ ,\quad a_f=0.99\ ,\quad a_{V'}=0.21\ ,\ \quad S=0.04\ ,\ T=0.07\ ;\nonumber\\ 
\chi^2_{\min}/\textrm{d.o.f.}&=&0.83\ ,\quad P\left(\chi^2>\chi_{\min}^2\right)=65\,\%\ ,\quad \textrm{d.o.f.}=15\ ,
\eea
where the number of degrees of freedom (d.o.f.) has decreased by one, since $a_{V'}$ is a free parameter. It is interesting to notice that the optimal value of $a_{V'}$ above is equal to the average $a_{V'}$, calculated over the 1000 data points, while the average values of $a_f$ and $a_V$ are, respectively, 0.98 and 1.03, which are very close to the corresponding optimal values given above. This shows that strong dynamics, that we used to determine the scanned range of values of the free parameters, generates rather naturally the coupling strengths favored by LHC data, at least once the direct constraints on the mass spectrum and the EW precision parameters are satisfied.

The 3MWT result in Eq.~\eqref{minchi2Vp} should be compared to the SM one:
\be\label{SMfitST}
\chi^2_{\min}/{\textrm{d.o.f.}}=0.89\ ,\quad P\left(\chi^2>\chi_{\min}^2\right)=60\%\ ,\quad {\textrm{d.o.f.}}=19\ .
\ee
While the SM fit is less satisfactory than the 3MWT one, it clearly shows that the SM is still perfectly viable in light of present collider data. It is instructive to notice that the fit performed with completely free coupling coefficients, therefore not motivated by any specific underlying theory, produces a worse fit than the 3MWT:
\bea\label{opta}
a_V&=&0.97^{+0.10}_{-0.11}\ ,\quad a_f=1.02^{+0.25}_{-0.32}\ , \quad  a_{V'}=0.21^{+0.16}_{-0.18}\ ,\nonumber\\
\chi^2_{\textrm{min}}/{\textrm{d.o.f.}}&=&0.85\ ,\quad P\left(\chi^2>\chi_{\textrm{min}}^2\right)=62\%\ ,\quad {\textrm{d.o.f}}.=14 \ .
\eea
This is because the underlying strong dynamics introduces a large correlation between $a_f$ and $a_V$, hence increasing the number of d.o.f by one, while achieving a $\chi^2_{min}$ very close to the corresponding result obtained with free coupling coefficients.

\section{Conclusions}\label{conc}

We have constructed a supersymmetric TC model and studied its low energy phenomenology in detail. We started from the ultraviolet complete model obtained by coupling Minimal Walking Technicolor and the SM with two Higgs doublets and supersymmetrizing the entire theory. We then considered a situation where the supersymmetry breaking occurs at scales much higher than the electroweak scale, leading to decoupling of all elementray scalars, gauginos, and higgsinos. What remains at low energies is a TC theory with global SU(3) chiral symmetry and its coupling with the SM fermions. We labeled this model 3MWT. 

We constructed the low energy effective Lagrangian of 3MWT and analysed its phenomenological consequences. We discussed the structure of the vacuum expectation values and emphasised the differences with respect to the related MWT model. Then we discussed how the observed light Higgs boson emerges from 3MWT. We computed the oblique corrections and performed a scan over the parameter space of the model to contrast it with the present data. The goodness of fit analysis, which we performed by using the observed Higgs coupling strengths as well as the $S$ and $T$ EW parameters, demonstrated that 3MWT yields a slightly better fit to the data than the SM. In particular, the role played by heavy composite vector resonances is critical, as their contribution to the diphoton decay of the light Higgs is absolutely necessary to bring the corresponding 3MWT prediction within the experimental constraints.  These composite vector resonances, having mass of ${\cal O}({\rm TeV})$, should be in principle observable at LHC, for example through Drell-Yan processes with two leptons in the final state or through a $W'$ decaying into three leptons plus missing energy.

\begin{acknowledgements} We thank R. Foadi for providing the code to evaluate the EW oblique corrections and for discussions. 

\end{acknowledgements}


\appendix

\section{EW Symmetry Breaking in Global SU$(3)$ Invariant Technicolor}
\label{SO3EWbreak}

The SU$(3)$ symmetry of the microscopic TC Lagrangian is spontaneously broken to the maximal diagonal subgroup, SO$(3)$. The symmetry breaking pattern leaves us with five broken  generators with associated Goldstone bosons. Such a breaking is driven by the condensate \be \langle \eta_i^\alpha \eta_j^\beta
\epsilon_{\alpha \beta} E^{ij} \rangle =\langle 2 U^\dagger_R U_L
+ D_L D_L\rangle \ , \label{conde}
 \ee
where the indices $i,j=1,\ldots,3$ denote the components
of the triplet of $\eta$, and the Greek indices indicate the ordinary
spin. The matrix $E$ is a $3\times 3$ matrix defined as
 \be E=\left(
\begin{array}{ccc}
0 & 0 & 1 \\
0 & 1 & 0 \\
1 & 0 & 0
\end{array}
\right) \ . \ee
The above condensate is invariant under an SO$(3)$ symmetry. It is convenient to separate the eight generators of SU(3) into the three that leave the vacuum invariant, $S^a$, and the remaining five that do not, $X^a$. Then the $S^a$ generators of the SO(3) subgroup satisfy the relation
\be
S^a\,E + E\,{S^a}^{T} = 0 \ ,\qquad {\rm with}\qquad  a=1,\ldots  ,  3 \ ,
\ee
so that ${\rm u}E{\rm u}^T=E$, for ${\rm u}\in$ SO(3). An explicit realization of the generators is shown in  Appendix \eqref{generators}. 

The scalar and pseudoscalar degrees of freedom, necessary to model the Goldstone bosons and spontaneous symmetry breaking, consist of a composite Higgs and its pseudoscalar partner, as well as five pseudoscalar Goldstone bosons and their scalar partners. These
can be assembled in the matrix
\be
M = \left[\frac{\sigma+i{\Theta}}{\sqrt{3}} I_3 + \sqrt{2}(i\Pi^a+\widetilde{\Pi}^a)\,X^a\right]E \ ,
\label{M}
\ee
which transforms under the full SU(3) group according to
\be
M\rightarrow {\rm u}M{\rm u}^T \ , \qquad {\rm with} \qquad {\rm u}\in {\rm SU(3)} \ .
\ee
The $X^a$'s, $a=1,\ldots,5$ are the generators of the SU(3) group which do not leave  the vacuum expectation value (VEV) of $M$ invariant
\be
\langle M \rangle = \frac{v}{\sqrt{3}}E \ .
\ee

\section{SU$(3)$ Generators\label{generators}}

The generators $S^i$ of SO$(3)$ satisfy $S^i E+ES^{i T}=0$. The other generators of SU$(3)$ are written as $X^i$. The generators are normalized as
\bea
\tr{[S^i S^j]}=\delta^{ij}/2 \;&\; \tr{[X^i X^j]}=\delta^{ij}/2 \;&\; \tr{[X^i S^j]}=0
\eea
and given in terms of the Gell-Mann matrices $\lambda^i$ by
\bea
S^1&=&\frac{1}{2\sqrt{2}}\left(\lambda ^1 - \lambda ^6\right)\\
S^2&=&\frac{1}{2\sqrt{2}}\left(\lambda ^2 - \lambda ^7\right)\\
S^3&=&\frac{1}{4}\left( \lambda ^3+\sqrt{3}\lambda ^8\right)\\
X^1&=&\frac{1}{2\sqrt{2}}\left(\lambda ^1 + \lambda ^6\right)\\
X^2&=&\frac{1}{2\sqrt{2}}\left(\lambda ^2 + \lambda ^7\right)\\
X^3&=&\frac{1}{4}\left(\sqrt{3}\lambda ^3 - \lambda ^8\right)\\
X^4&=&\frac{1}{2}\lambda ^4\\
X^5&=&\frac{1}{2}\lambda ^5
\eea
Using the generators above, it is straightforward to obtain the 
vector and axial-vector charge eigenstates and their elementary particle content 
from Eqs.~(\ref{Vmtx},\ref{Vcont}). First note that the charge operator is $Q=S^3$. We find first the linear
combinations of the generators corresponding to charge eigenvalues $0$, $\pm 1$ and $\pm 2$. Then we project the corresponding vector states, e.g. $v^0_\mu=2 {\rm{Tr}}(S^3A_\mu)$, and obtain:
\bea
v^{0}_\mu&\equiv& \frac{A^{3}_\mu+\sqrt{3}A^{ 8}_\mu}{2}\sim\bar{U}_L\gamma_\mu U_L+\bar{U}_R\gamma_\mu U_R\ ,\nonumber\\
v^{+}_\mu&\equiv& \frac{A^{1}_\mu-A^{6}_\mu}{2}-i\frac{A^{2}_\mu-A^{7}_\mu}{2}\sim\bar{D}_L\gamma_\mu U_L+\bar{D}^{\rm c}_L\gamma_\mu U_R\ ,\nonumber\\
v^{-}_\mu&\equiv& \frac{A^{1}_\mu-A^{6}_\mu}{2}+i\frac{A^{2}_\mu-A^{7}_\mu}{2}\sim\bar{U}_L\gamma_\mu D_L+\bar{U}_R\gamma_\mu D^{\rm c}_L\ ,\nonumber\\
a^{0}_\mu&\equiv& \frac{\sqrt{3} A^{3}_\mu-A^{8}_\mu}{2}\sim\bar{U}_L\gamma_\mu U_L-\bar{U}_R\gamma_\mu U_R-2 \bar{D}_L\gamma_\mu D_L\ ,\nonumber\\
a^{+}_\mu&\equiv& \frac{A^{1}_\mu+A^{6}_\mu}{2}-i\frac{A^{2}_\mu+A^{7}_\mu}{2}\sim\bar{D}_L\gamma_\mu U_L-\bar{D}^{\rm c}_L\gamma_\mu U_R\ ,\nonumber\\
a^{-}_\mu&\equiv& \frac{A^{1}_\mu+A^{6}_\mu}{2}+i\frac{A^{2}_\mu+A^{7}_\mu}{2}\sim\bar{U}_L\gamma_\mu D_L-\bar{U}_R\gamma_\mu D^{\rm c}_L\ ,\nonumber\\
\Omega^{++}_\mu&\equiv& \frac{A^{4}_\mu-i A^{5}_\mu}{\sqrt{2}}\sim \bar{U}^{\rm c}_R \gamma_\mu U_L\ ,\quad
\Omega^{--}_\mu\equiv \frac{A^{4}_\mu+i A^{5}_\mu}{\sqrt{2}}\sim \bar{U}_L \gamma_\mu U^{\rm c}_R\ .
\label{Vcontent}\eea
The particle contents given above reproduce the corresponding results in \cite{Foadi:2007ue} if one 
applies there the substitution $D_R\rightarrow D_L^{\rm c}$.

\section{Squared Mass Matrices}\label{sMMx}

For the neutral scalar and pseudoscalar states, the charged and doubly charged states, the squared mass matrices are, respectively
\be\label{Mhn}
{\cal M}^2_{\bar{h}^0}=\left(
\begin{array}{cc}
 2 v_{\sigma }^2 \left(\lambda +2 \lambda '\right) & 2 v_{\Delta } v_{\sigma } \left(\lambda -2 \lambda''\right) \\
 2 v_{\Delta } v_{\sigma } \left(\lambda -2 \lambda''\right) & 2 \left(v_{\sigma }^2 \lambda''+v_{\Delta }^2 \left(\lambda +4 \lambda '\right)\right)
\end{array}
\right)\ ,
\ee
in the $\Re\left(\sigma^0\right),\Re\left(\Delta^0\right)$ basis,
\be
{\cal M}^2_{\bar{\pi}^0}=\left(
\begin{array}{cc}
 8 v_{\Delta }^2 \lambda'' & 4 v_{\Delta } v_{\sigma } \lambda'' \\
 4 v_{\Delta } v_{\sigma } \lambda'' & 2 v_{\sigma }^2 \lambda''
\end{array}
\right)\ ,
\ee
in the $\Im\left(\sigma^0\right),\Im\left(\Delta^0\right)$ basis,
\be
{\cal M}^2_{\bar{h}^\pm}=\left(
\begin{array}{cc}
 2 v_{\sigma }^2 \left(\lambda''+\lambda '\right) & 2 \sqrt{2} v_{\Delta } v_{\sigma } \left(\lambda''+\lambda '\right) \\
 2 \sqrt{2} v_{\Delta } v_{\sigma } \left(\lambda''+\lambda '\right) & 4 v_{\Delta }^2 \left(\lambda''+\lambda '\right)
\end{array}
\right)\ ,
\ee
in the $\Delta^\pm,\sigma^\pm$ basis,
\be
{\cal M}^2_{\bar{h}^{\pm\pm}}=\left(
\begin{array}{cc}
 2 v_{\sigma }^2 \lambda''-4 \left(v_{\Delta }^2-v_{\sigma }^2\right) \lambda ' & 4 v_{\Delta }^2 \lambda''+2 v_{\sigma }^2 \lambda ' \\
 4 v_{\Delta }^2 \lambda''+2 v_{\sigma }^2 \lambda ' & 2 v_{\sigma }^2 \lambda''-4 \left(v_{\Delta }^2-v_{\sigma }^2\right) \lambda '
\end{array}
\right)\ ,
\ee
in the $\Delta^{\pm\pm},\delta^{\pm\pm}$ basis.

We define, besides $\epsilon$ in Eq.~\eqref{Cmu}, the following dimensionless parameters:
\be
x=\frac{g_L v_w}{2 m_A}\ ,\quad t_\rho=\frac{\sqrt{2} v_\Delta}{v_\sigma}\ ,\quad z_i= \left(\frac{g_{\rm TC} v_w}{2 m_A}\right)^2 r_i\ ,\ i=1,2,3\ .
\ee
Then the non-zero terms of the charged vector boson squared mass matrix (which by definition is symmetric) are
\bea\label{sqMVc}
\left({\cal M}_{\bar{W}}^2\right)_{1,1}&=&m_A^2 \left[x^2+\epsilon ^2  \left(1+z_1-\frac{z_3}{2} \left(1+c_{\rho }^2\right) \right)\right]\ ,\nonumber\\ 
   \left({\cal M}_{\bar{W}}^2\right)_{2,2}&=&m_A^2 \left[1+z_1+z_2 s_{2 \rho }-\frac{z_3}{2} \left(1+c_{\rho }^2\right)\right]\ ,\quad 
\left({\cal M}_{\bar{W}}^2\right)_{1,2}=-\frac{\epsilon}{\sqrt{2}}    \left({\cal M}_{\bar{W}}^2\right)_{2,2}\ ,\nonumber\\ 
   \left({\cal M}_{\bar{W}}^2\right)_{3,3}&=&m_A^2 \left[1+z_1-z_2 s_{2 \rho }-\frac{z_3}{2} \left(1+c_{\rho }^2\right)\right]\ ,\quad
\left({\cal M}_{\bar{W}}^2\right)_{1,3}=-\frac{\epsilon}{\sqrt{2}} \left({\cal M}_{\bar{W}}^2\right)_{3,3}\ ,
\eea
in the $\tilde{W}_\mu^\pm,V_\mu^\pm,A_\mu^\pm$ basis, with furthermore the squared mass of the doubly charged vector boson given by
\be\label{smO}
m^2_{\Omega}=m_A^2 \left[ 1+2 c_{\rho }^2 \left(z_1+z_2\right)-\frac{z_3}{2} \left(1+c_{\rho }^2\right) \right]\ .
\ee
Finally, the non-zero terms of the neutral vector boson squared mass matrix in the $\tilde{W}_\mu^3,B_\mu,V_\mu^3,A_\mu^3$ basis  are
\bea
\left({\cal M}_{\bar{Z}}^2\right)_{1,1}&=&m_A^2 \left[x^2 \left(1+s_{\rho }^2\right)+\epsilon ^2 \left(1+z_1-\frac{1}{2} \left(1+c_{\rho }^2\right) z_3+z_2 s_{\rho }^2\right)\right]\ ,\nonumber\\ 
\left({\cal M}_{\bar{Z}}^2\right)_{1,2}&=&-m_A^2 \left[x^2 \left(1+s_{\rho }^2\right)+\epsilon ^2 \left(z_1-c_{\rho }^2 \left(2 z_1-z_2\right)+z_2\right)\right] t_{\xi}\ ,\nonumber\\ 
\left({\cal M}_{\bar{Z}}^2\right)_{2,2}&=&m_A^2 \left[x^2 \left(1+s_{\rho }^2\right)+\epsilon ^2 \left(3+z_1+c_{\rho}^2 \left(4 z_1-5 z_2\right)+z_2-\frac{3}{2} \left(1+c_{\rho }^2\right) z_3\right)\right]t_{\xi }^2\ ,\nonumber\\ 
   \left({\cal M}_{\bar{Z}}^2\right)_{3,3}&=&m_A^2 \left[1+2 c_{\rho}^2 \left(z_1-z_2\right)-\frac{z_3}{2} \left(1+c_{\rho }^2\right) \right]\ ,\nonumber\\ 
\left({\cal M}_{\bar{Z}}^2\right)_{1,3}&=&-\frac{\epsilon}{2}    \left({\cal M}_{\bar{Z}}^2\right)_{3,3}\ ,\quad
 \left({\cal M}_{\bar{Z}}^2\right)_{2,3}=-\frac{3}{2} \epsilon\, t_{\xi}   \left({\cal M}_{\bar{Z}}^2\right)_{3,3}\ ,\nonumber\\ 
   \left({\cal M}_{\bar{Z}}^2\right)_{4,4}&=&m_A^2 \left[1+z_1-z_2 s_{2 \rho }-\frac{z_3}{2} \left(1+c_{\rho }^2\right)\right]\ ,\nonumber\\
\left({\cal M}_{\bar{Z}}^2\right)_{1,4}&=&-\frac{\sqrt{3}}{2}\epsilon \left({\cal M}_{\bar{Z}}^2\right)_{4,4}\ ,\quad
 \left({\cal M}_{\bar{Z}}^2\right)_{2,4}=\frac{\sqrt{3}}{2} \epsilon\, t_\xi    \left({\cal M}_{\bar{Z}}^2\right)_{4,4}\ .
\eea

\section{$S$ and $T$ Parameters for General Neutrino Mass Matrix}\label{STgen}

The most general mass terms for a pair of right- and left-handed neutrinos is defined by
\begin{eqnarray}  
{\cal L}\supset-m_E \bar{E}_R E_L-\frac{1}{2}n_L^T Mn_L+{\rm{h.c.}},~~~M=\left(\begin{array}{cc} M_L & m_D \\ m_D & M_R\end{array}\right),\ n_L=(N_{L}, \bar{N}_{R})^T
\end{eqnarray}
 with 
 eigenvalues
\begin{eqnarray}
\lambda_{1,2}=\frac{1}{2}\left[(M_L+M_R)\pm\sqrt{(M_L-M_R)^2+4m_D^2}\right]\ .
\end{eqnarray}
The contributions of the corresponding heavy neutrinos mass eigenstates and of the heavy electron $E$ to the $S$ and $T$ parameters have been derived in terms of integral functions in \cite{Antipin:2009ks}. From those, we derived the corresponding explicit results: 
\bea
S &=&\frac{1}{12 \pi }\left[1+2 c_{\zeta }^4 \left(1+\log\nu _1^2\right)-2 \log\nu _E^2+2 s_{\zeta }^4 \left(1+\log\nu _2^2\right)\right]\nonumber\\
&+&\frac{s_{2 \zeta }^2}{36 \pi}
   \frac{9 \left(1-\log\nu _1^2\right) \nu _1^4 \nu _2^2-9 \left(1-\log\nu _2^2\right) \nu _1^2 \nu _2^4-\left(1-3 \log\nu _1^2\right) \nu
   _1^6+\left(1-3 \log\nu _2^2\right) \nu _2^6}{  \left(\nu _1^2-\nu _2^2\right)^3}\nonumber\\
&-&\frac{(-1)^{\beta } s_{2 \zeta }^2}{8 \pi} \frac{\nu _1 \nu _2
   \left(\nu _1^4-2 \nu _1^2 \nu _2^2 \log\frac{\nu _1^2}{\nu _2^2}-\nu _2^4\right)}{\left(\nu _1^2-\nu _2^2\right)^3}\ ,
   \eea
\bea\label{genT}
T &=& \frac{\Lambda_{NP}^2}{64 \pi c^2_{\xi} s^2_{\xi} m_Z^2}\left[\vphantom{\frac{\left(1-2  \log \nu _1^2\right) \nu _1^4}{\nu _1^2-\nu _2^2}}16 c_{\zeta }^4  \nu _1^2 \log \nu _1^2+16 s_{\zeta }^4  \nu _2^2\log \nu _2^2+8 \nu _E^2 \log \nu _E^2\right.\nonumber\\
&-&s_{2 \zeta }^2 \frac{\left(1-2
   \log \nu _1^2\right) \nu _1^4-\left(1-2 \log \nu _2^2\right) \nu _2^4}{\nu _1^2-\nu _2^2}+4 (-1)^{\beta } s_{2 \zeta }^2 \frac{ \left(1-\log \nu _1^2\right) \nu _1^3 \nu
   _2-\left(1-\log \nu _2^2\right) \nu _1 \nu _2^3}{\nu _1^2-\nu _2^2}\nonumber\\
   &+&4 c_{\zeta }^2 \frac{\left(1-2 \log \nu _1^2\right) \nu
   _1^4-\left(1-2 \log \nu _E^2\right) \nu _E^4}{\nu _1^2-\nu _E^2} + \left.4 s_{\zeta }^2 \frac{\left(1-2 \log \nu _2^2\right) \nu _2^4-\left(1-2 \log\nu_E^2\right) \nu _E^4}{\nu _2^2-\nu _E^2}\right]\ ,
\eea
 where $\Lambda_{NP}$ is the given renormalization scale, $\xi$ is the EW mixing angle, and
 \be
 \nu_1=\frac{\lambda_1}{\Lambda_{NP}}\ ,\quad \nu_2=\frac{\lambda_2}{\Lambda_{NP}}\ ,\quad \nu_E=\frac{m_E}{\Lambda_{NP}}\ ,\quad t_{2\zeta}=\frac{2m_D}{M_R-M_L}\ ,\quad \beta=\frac{1}{2}\left[1+\left(\sqrt{\frac{\lambda_1}{|\lambda_1|}}^{~\ast}\sqrt{\frac{\lambda_2}{|\lambda_2|}}\right)^2\right]\ .
\ee
 In the limit $M_R\rightarrow \infty$, and $M_L=m_E\equiv m_U$, one recovers the results in Eqs.~\eqref{STMaj}.
 
 \bibliography{3MWT.bib}
 
 \end{document}